\documentclass[12pt,floatfix,nofootinbib,aps]{revtex4-1}
\usepackage{graphicx}
\usepackage{color}
\usepackage{adjustbox}
\usepackage{multirow}
\usepackage{ulem} 
\usepackage{amsmath}
\usepackage[font=footnotesize,labelfont=rm]{caption}
\renewcommand{\thesubsection}{\Roman{section}.\arabic{subsection}}
\makeatletter
\renewcommand{\p@subsection}{\Roman{section}.\arabic{subsection}\expandafter\@gobble}
\renewcommand{\p@subsubsection}{\thesubsection.\arabic{subsubsection}\expandafter\@gobble}
\makeatother

\begin{document}
\title{Pygmy resonances: what's in a name?}
\author{R.A. Broglia$^{a,b} $, F. Barranco$^c$, A. Idini$^d$, G. Potel$^e$, E. Vigezzi$^f$}
\affiliation{
$^a$ Dipartimento di Fisica, Universit\`a di Milano, Via Celoria 16, 20133 Milano, Italy. \\  
$^b$ The Niels Bohr Institute, University of Copenhagen, Blegdamsvej 17, Copenhagen, Denmark.\\  
$^c$ Departamento de Fisica Aplicada III, Escuela Superior de Ingenieros, Universidad de Sevilla, 
   Camino de los Descubrimientos s/n,  
   41092 Sevilla, Spain.\\
$^d$ Division of Mathematical Physics, Department of Physics, LTH, Lund University, Post Office Box 118, S-22100 Lund, Sweden\\
$^e$ National Superconducting Cyclotron Laboratory,  Michigan  State University,  
   East Lansing,  Michigan  48824,  USA\\
$^f$ INFN, Sezione di Milano, Via Celoria 16, 20133 Milano, Italy.\\
}
\date{\today}

\begin{abstract}
The centroid, width and percentage of energy weighted sum rule of dipole resonances can be strongly affected by dynamical fluctuations and static deformations of the nuclear surface, deformations and fluctuations which, in turn, depend on pairing, and thus on Cooper pairs. 
Because of angular momentum conservation, such insight is restricted, to lowest order,  to deformations of quadrupole and monopole type. The latter being closely connected with the neutron (excess) skin and thus with soft dipole modes.
From the values $(N-Z)/A \approx 0.18$, 0.21, and 0.45 for the nuclei $^{122}$Sn, $^{208}$Pb, and $^{11}$Li, it is expected that  the latter system, 
which is weakly bound  by pairing effects  (spatially extended single Cooper pair and odd proton acting as spectator), constitutes an attractive laboratory to study the properties of soft $E1$--modes and thus of isospin nuclear deformation. From the calculation of the full dipole response function in QRPA, discretizing the continuum in a spherical box of radius of 40 fm, one finds a GDR with centroid $E_x \approx 24$ MeV, width $\Gamma \approx$ 11 MeV and carrying 90\% of the EWSR, and a low--lying collective resonance characterized by $E_X = 0.75$ MeV, $\Gamma = 0.5$ MeV and $6.2\%$ EWSR  
The wave function of the latter resonance is built out of about fifteen components (both protons and neutrons), typical of a collective mode. The transition densities indicate this soft $E1$--mode to be generated by surface density oscillation of the neutron skin ($\Delta r_{np} \approx 1.71$ fm) relative to an approximately isospin--saturated core. 
Through a detailed study of the full dipole response of $^{11}$Li we will draw a comparison between the soft $E1$--mode of this halo nucleus and the PDR of heavy stable nuclei, pointing to the physical similarities and also to the basic differences.
\end{abstract}

\maketitle
% \setstretch{2}
% \tableofcontents
\section{Introduction}
\label{sect:intro}
Dipole modes are affected by deformations. In particular quadrupole deformations of the nuclear surface (measured by $\beta_2$). But also deformations taking place in other, more abstract, spaces like gauge (pairing, measured by the number of Cooper pairs $\alpha_0$) and isospace. This last case being closely connected with the neutron skin (measured by $\Delta r_{np} = \langle r^2 \rangle^{1/2}_n - \langle r^2 \rangle^{1/2}_p$), and the appearance of soft $E1$--modes.

The main purpose of the present  paper, is that of using the properties of the soft $E1$--mode of $^{11}$Li in trying to achieve a quantitative understanding of the mechanism at the basis of
the origin and dynamics of the neutron skin of this nucleus, and thus of the  associated low--lying dipole mode. This aim is similar to that which saw the nuclear physics community \cite{Barranco:01, Brida:06, Lenske:01, Tanihata:08, Potel:10} provide a consistent picture of the many--body aspects of the $^{11}$Li ground state.
That is, the incipient, dynamical deformation in gauge space, at the level of single Cooper pair.
From the prediction of the microscopic structure and thus of e.g. the (p,t) absolute   differential cross sections \cite{Barranco:01, Brida:06, Lenske:01}, to experimental confirmation \cite{Tanihata:08} (see also \cite{Potel:10}). A result made possible by the remarkable technical developments which led to inverse kinematics and active targets. These techniques and devices, coupled to experimental ingenuity, allowed for the detection of single Cooper tunneling to selected quantum states, involving the ms lifetime exotic nucleus $^{11}$Li. A feat which we, ever so briefly, attempt to set in connection with similar breakthroughs in quantum materials (single electron devices), and quantum optics (tweezers, able to handle single biological molecules), in trying to convey our remarkable nuclear physics field to a broader audience.

In the case of the $E1$--soft mode of $^{11}$Li, pairing is not only important, but it is mainly governed by the renormalization of the bare $^{1}$S$_0$ NN--force resulting from the exchange of this dipole mode between the least bound neutrons (bootstrap mechanism). One is thus confronted with a close, essentially symbiotic, relation between deformation in isospace ($\Delta r_{np} \approx 1.7$ fm) and in gauge space ($\alpha_0 = 1$). Within this context it can be  posited  that, were    the $E1$--soft mode of $^{11}$Li not a collective mode, $^{11}$Li  would not be bound. A subject which is further discussed in Sect. \ref{sec:introduction} (in particular in connection with the work of Lenske et al. \cite{Lenske:01}). In that section we also dwell upon the subtle interplay between deformation (spontaneously broken symmetry) in different spaces, in particular gauge and isospin spaces. We conclude Sect. \ref{sec:introduction}  by reminding that important advances in the understanding of the properties of collective motion in nuclei, but also in the study of more complex 
systems, have been obtained by measurements of vibrations or other periodic  motions, revealing the anisotropy of the local (mean) field. Measurements which in quadrupole deformed nuclei are geometrically parametrized by the radius $R = R_0 (1 + \beta_2 Y_{20}(\theta))$. Parametrization which can be heuristically generalized, in the case of distortion in isospace through $R(^{11}$Li$) = R_0 (^{11}$Li$) ( 1 + \beta_0 Y_{00})$ (Sect. \ref{sec:microscopic}). This is mainly done for the purpose of attempting at making the analogy clear to a broad audience of physics practitioners. The short remarks referred to NMR attempt at doing the same in connection with the broader scientific community, in particular biophysicists.

In Section \ref{sec:dipole} the mechanism for generating a low--energy dipole state carrying few percent of the TRK sum rule, and thus displaying a significant amount of collectivity is discussed. 
At this point, two comments are in place.  The first one, regards the question of the so called continuum effect \cite{Catara:96}, implying by this the fact that the occurrence of weakly bound states towards the neutron drip line gives rise to low--lying concentration of 
dipole strength in the continuum. A result we find in harmony with the outcome  of our calculations, as can be seen from
Fig. 4,  where the  corresponding uncorrelated neutron particle-hole  dipole response is compared with that  of \cite{Catara:96}, in keeping with the fact that in this reference a square well, independent--particle model is used.
The second, closely connected with the first one,  
concerns the need to treat simultaneously the full dipole response, i.e. both soft $E1$--modes and the GDR, so as to ensure the center of mass of the system to be at rest and thus conservation of the TRK sum rule.
And, as a bonus, to have the information  needed to assess whether the low-energy state is a collective, resonant mode or less.
%A valid statement concerning the unperturbed response function, i.e. uncorrelated particle--hole excitations (as can be seen from Fig. \ref{fig:4} where our neutron unperturbed function is compared with that of \cite{Catara:96}), in keeping with the fact that in \cite{Catara:96} a square well, independent--particle model of like {\color{red} particles} is used. 
%However, the central conclusion of the paper, namely: ``An interesting feature of the novel {\color{red} structures} is that they are not born at the expense of the ordinary strength at higher excitation energies'' is not applicable, in any case not concerning the dipole response function, in keeping with the fact that the TRK energy weighted sum rule depends only on the number of particles.
 Within this context, and because the TRK dipole EWSR only depends, as a variable, on the number of particles, whatever dipole 
 strength is found in the soft E1-mode, is at the expenses of the GDR.
  A question discussed in detail in section \ref{subsec:e1}.

\begin{figure}[ht!]
	\begin{center}
		\includegraphics[width=0.49\textwidth]{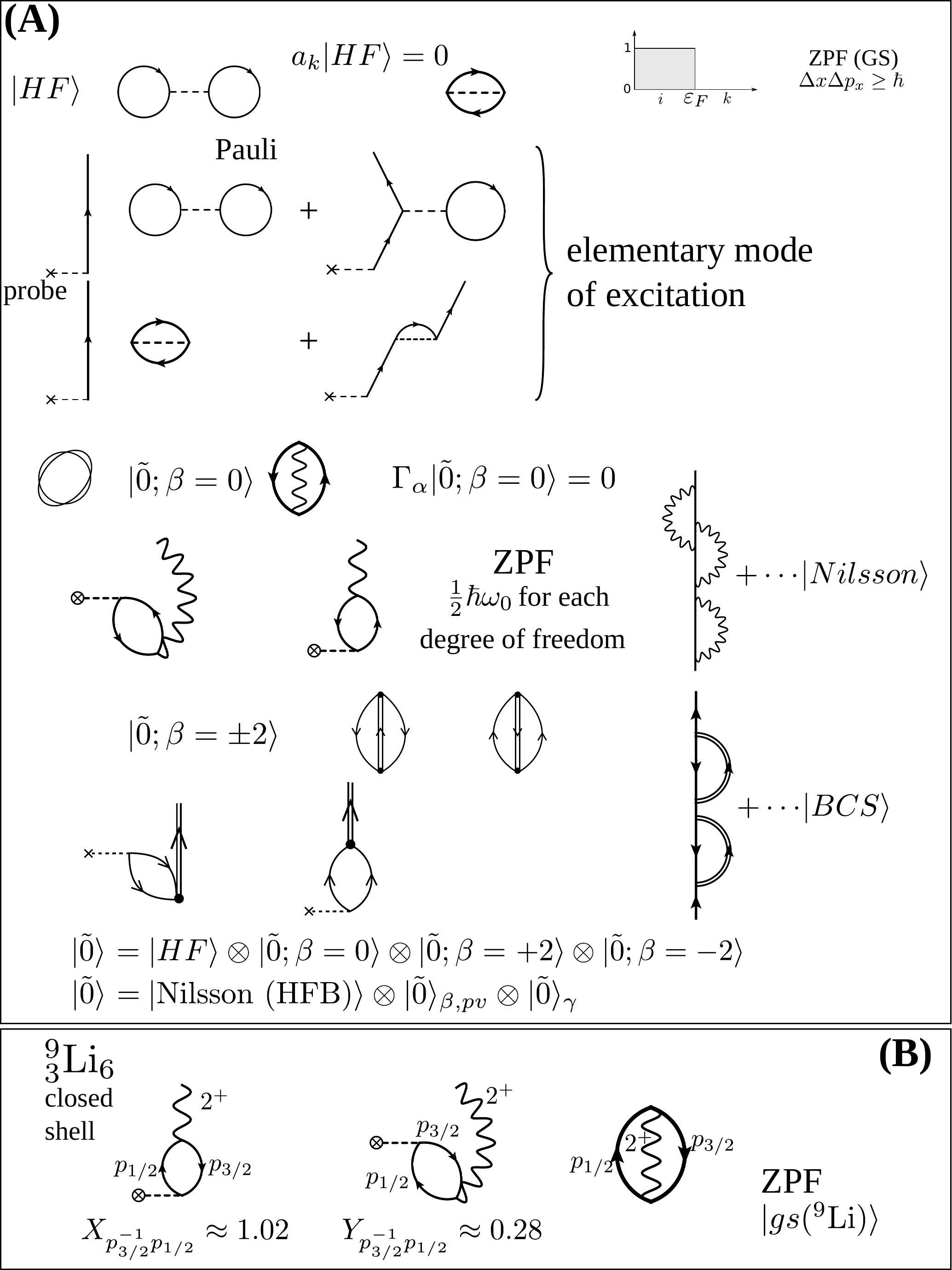}
		\includegraphics[width=0.49\textwidth]{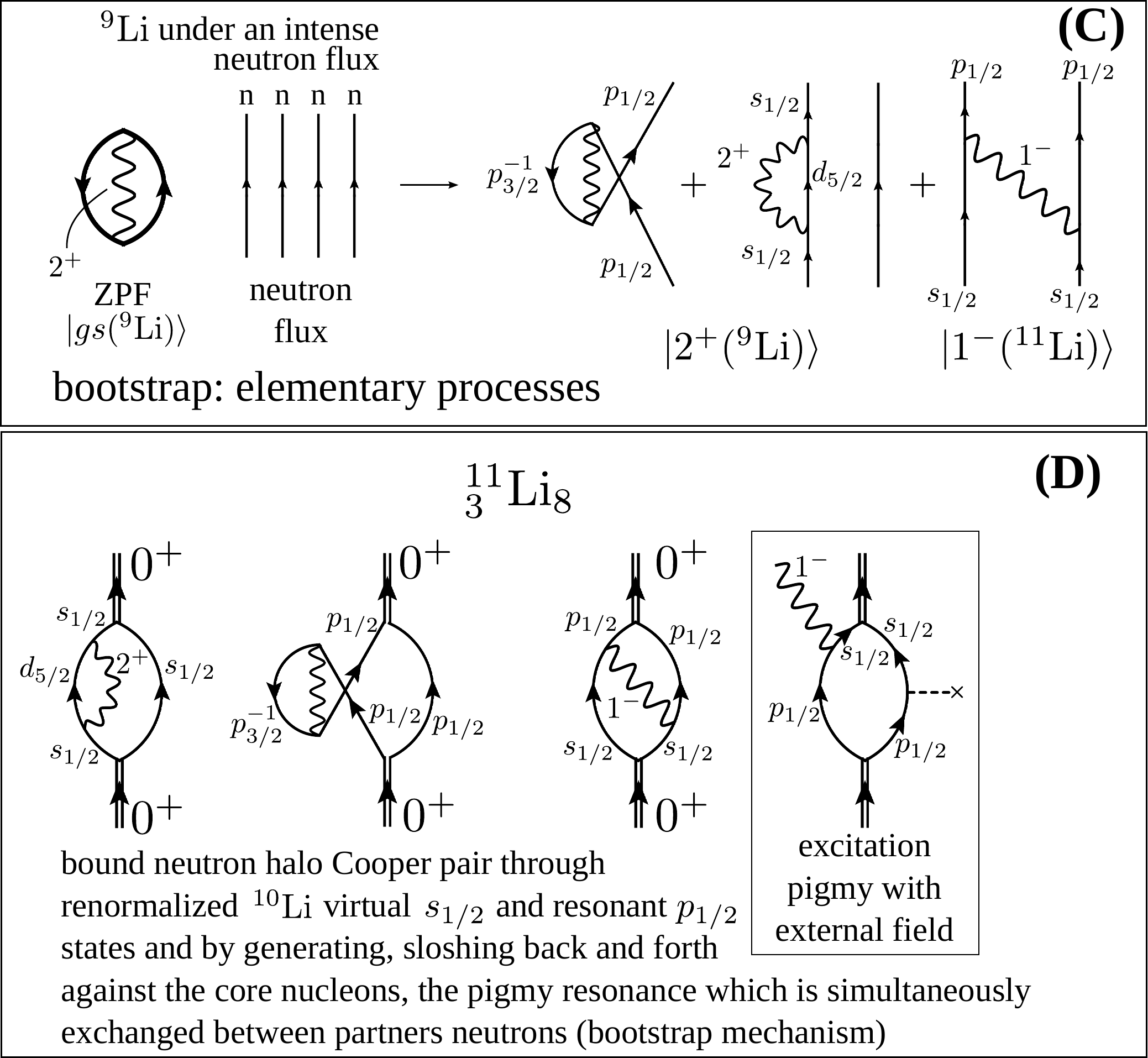}
	\end{center}
	\caption{ Each elementary mode of excitation has its proper ground state, tantamount to saying its individual quantal ZPF, as well as to refer to the possibility of exciting each mode making use of the specific probe, as can be seen in \textbf{(A)}: (upper) independent-particle motion and one-particle transfer $(\beta = \pm 1)$; (middle) surface, particle-hole-like ($\beta = 0$) vibrations; (lower) pairing vibrations and two-particle transfer $(\beta = \pm 2)$. \textbf{(B)} Quadrupole excitations (RPA) and ZPF of the quadrupole mode of the core of $^{11}$Li (of notice that the $p_{3/2}(\pi)$ proton single-particle state is not explicitly shown, being considered as a spectator).\textbf{(C)} the lowest $2^+$ vibration of $^9$Li evidenced also through the ZPF become real in a \textit{Gedanken experiment} in which a target of $^{9}$Li subject to an intense flux of neutrons, essentially traps two of them at threshold ($s_{1/2}$ and $p_{1/2}$). By sloshing back and forth against the protons and neutrons of the $^{9}$Li core, it generates a soft dipole resonance (labeled pygmy for convenience) which exchanged between the $s^2_{1/2} (0)$, $p^2_{1/2} (0)$ configurations (bootstrap mechanism), binds the Cooper pair neutron halo (neutron halo pair addition mode). \textbf{(D)}: in other words, the soft (dipole) resonance in $^{11}$Li cannot exist without the two halo neutrons, nor the two halo neutrons can bind the $^9$Li core without the soft $E1$--resonance, the two modes acting in a symbiotic fashion. }
\label{fig:1}
\end{figure}

In Section \ref{sec:elementary} we briefly discuss the theoretical basis to calculate the ground state (single Cooper pair) of $^{11}$Li (see also Fig. \ref{fig:1}) in terms of elementary modes of excitation, i.e. collective surface and pairing vibrations and single--particle motion, and their interweaving through the particle--vibration coupling vertex.

A discussion of the microscopic calculations of the $E1$--soft mode of $^{11}$Li and of the results is provided in Sect. \ref{sec:microscopic}. In connection with the calculation of the ground state of $^{11}$Li (Eqs. (\ref{eq:gs})-(\ref{eq:onu}), see also Fig. \ref{fig:1} (D)) we are confronted with $^{1}$S$_0$ (singlet even, SE) pairing correlations at very low densities, as expressed by Eq. (\ref{eq:dens}). And because of the poor overlap between neutrons belonging to the neutron skin (halo) and to the core, a screened NN--pairing interaction.
Within this context, the paper of Lenske et al. \cite{Lenske:01} constitutes an important contribution to the subject in which the ground state of $^{11}$Li is calculated, continuum effects being exactly described when solving the Gorkov--equations, making use of SE Brueckner G--matrix in the particle--particle channel.

In subsection \ref{subsec:e1} the full dipole response function is presented, as well as the wavefunctions and transition densities,  testifying to the collectivity of the soft $E1$--mode of $^{11}$Li. The associated transition densities compare at profit with that of the PDR of $^{208}$Pb \cite{Ryezayeva:02} (see also \cite{Lane:71}) and of $^{122}$Sn \cite{Tsoneva:08}. From these results, as well as the overall account of the experimental findings (in particular those of ref \cite{Kanungo:15}), one can posit that the calculated value of $\Delta r_{np}$ represents a quantitative  estimate of the neutron skin of $^{11}$Li.

Section \ref{sec:hindsight} provides hindsight regarding the physical stability and soundness of the results, while the conclusions are collected in Sect. \ref{sec:conclusions}.

There are two appendices. In the first, a simple estimate of the overlap between halo and core single--particle wavefunctions is provided, making use of familiar concepts of superconductivity in metals and associated BCS description, namely the coherence length. In appendix B technical details concerning the treatment of the continuum are given.

\section{Soft E1--modes and pairing}\label{sec:introduction}
Nuclei respond elastically to sudden solicitations and plastically to slowly changing fields. Systems which display both elastic and plastic behavior are well known from other fields of physics (non-Newtonian solids). In particular, the natural caoutchouc, the rubber out which car tires were made (see \cite{Bertsch:83a} last paragraph of third column p. 62; see also \cite{Rogers:18}). Rigidity in nuclei is provided by the shell structure, and  is associated to the energy difference between major shells, which in medium-heavy nuclei, amounts to $\hbar \omega_0 \approx 7-8$ MeV. This quantity fixes the characteristic time over which external fields have to vary to be able to excite elastic nuclear modes, i.e., $\tau = 1/\omega_0 \approx 10^{-22}$ s. When the atomic nucleus is subject to a field, either external or internal (e.g. zero point fluctuations), which changes slowly in time $(\Delta t \gg \tau)$ the system deforms plastically, and the single-particle levels undergo (dynamical, static) Jahn-Teller-like splitting \cite{Bohr:69a}.

Because pairing thrives on large degeneracies, much has been discussed concerning the competition between pairing and (quadrupole) deformation \cite{Bohr:75}. But pairing can tolerate some amount of degeneracy breaking, and nuclear superfluidity induced by pairing, plays an essential role in allowing for adiabatic, large amplitude surface modes of different multipolarities. Examples of this combined plastic behavior is provided by low-lying surface vibrations observed throughout the mass table and by exotic decay, in particular by the process $^{223}\textrm{Ra} \rightarrow ^{14}\textrm{C} + ^{209}\textrm{Pb}$, let alone nuclear fission (\cite{Barranco:88},  \cite{Brink:05} Ch. 7 and refs. therein; see also \cite{Bulgac:16},\cite{Bulgac:18},\cite{Grineviciute:18}). Within this broad scenario one can posit that \textit{the relationship between pairing and low-lying vibrational modes is symbiotic, not competitive}. 

Pairing reduces in a consistent way the inertia of low-lying nuclear vibrations. Thus, it increases nuclear collectivity \footnote{In the case of the exotic decay, $^{223}\textrm{Ra} \rightarrow ^{14}\textrm{C} + ^{209}\textrm{Pb}$, the inertia is changed from the value of $D/\hbar^2 = 1690\textrm{ MeV}^{-1}$, assuming $^{223}$Ra to be normal, to  $D/\hbar^2 = 29.1\textrm{ MeV}^{-1}$, by properly considering that this nucleus is superfluid. The associated exotic decay lifetimes are $\lambda = 2\times10^{-83}\textrm{ sec}^{-1}$ and $\lambda = 1.8\times10^{-16}\textrm{ sec}^{-1}$, respectively, the experimental value being $\lambda_{\textrm{exp}} = 4.3\times10^{-16}\textrm{ sec}^{-1}$ (\cite{Brink:05}, Ch. 7, p. 159).} as measured by the transition amplitude $\left(\hbar \omega/2 C\right)^{1/2} = \left(\hbar/2\sqrt{CD}\right)^{1/2}$ (\cite{Brink:05} Eq. (8.46) p. 190; see also Sect. 7.3 p. 165).  
Exchanged between nucleons moving in time reversal  states lying close to the Fermi energy,
these collective vibrations renormalize the bare pairing interaction by about a factor of 2 
 (\cite{Barranco:99,Terasaki:02b, Gori:05,Idini:12,Idini:16b}; see also \cite{Brink:05}, Ch. 10 and refs. therein). Within this context ---pairing fostering deformation--- we will show the important role that already a single Cooper pair can play.
In particular, in the case of the neutron halo pair addition mode of $^{9}$Li resulting in the $|gs(^{11}$Li)$\rangle)$ assuming the odd $p_{3/2}(\pi)$ proton to play the role of a spectator. In this case, the symbiosis also encompasses an antenna-like mode, namely the soft dipole mode, resulting in a system at the edge of acquiring a permanent (vortex-like) dipole moment (\cite{Avogadro:07}; see also \cite{Bertsch:94} p.41 and  \cite{Ryezayeva:02} last paragraph first column p. 272502-4).

Cooper pairs are the building blocks of fermionic superconductivity and superfluidity in condensed matter, $^3$He, finite nuclei, neutron stars, cold fermionic gases, etc \cite{Cooper:56,Schrieffer:64,Bardeen:57a,Bardeen:57b,Osheroff:72,Leggett:75,Wheatley:75,Vollhardt:90,Bohr:58,Combescot:08,Spuntarelli:10,Giorgetti:05,Cooper:11,Broglia:13book}.
The ability  to study the components of light, electric current and metabolism in terms of single-photon, single-electron and single-protein experiments (cf. e.g. \cite{Walborn:02,Postma:01,Shank:10,Caldarini:14} and refs. therein) has contributed much to the understanding of 
the (macroscopic)  phenomena of which these entities are the building blocks. Similarly, the manipulation and characterization of a
single Cooper pair of nucleons is important for the understanding of pairing in nuclei. 
Such a feat has been  accomplished in exotic halo nuclei, in the case in which 
a Cooper pair is barely bound to the Fermi surface of the core \cite{Tanihata:08,Potel:10}. 
The technique which allows to probe, through single-pair tunneling processes 
(see \cite{Potel:13} and refs. therein), the nuclear embodiment of 
 Cooper model is inverse kinematics  (see e.g. \cite{Tanihata:08,Tanihata:13,Tanihata:16}).

Van der Waals interaction plays a central role in the case of many-body proton 
and electron systems, from atoms to macromolecules (e.g. proteins). However, as a rule,
dipole zero point fluctuations (ZPF) do not contribute  to the renormalisation 
of the NN--bare interaction.
%are not significant to renormalize the overall low-energy structure properties of nuclei. 
Quadrupole and octupole nuclear vibrations are the lowest multipolarity modes 
leading to sizable contributions to the induced interaction  \cite{Bohr:75,Barranco:99,Brink:05,Terasaki:02b,Gori:05}.
Although one pays a price to separate both protons from electrons (Coulomb energy, plasmon) as well as neutron from protons (symmetry energy, Giant Dipole Resonances \cite{Bohr:75,Bertsch:94,Bortignon:98,Harakeh:02}), the relative masses of the pair of particles displaying an antenna-like motion are very different in the two cases ($m_e/M_p \approx 5 \times 10^{-4}$, $M_n/M_p \approx 1$).

Unexpectedly, the situation is quite different in exotic halo nuclei like $^{11}$Li, 
in which case a very low energy ($\lesssim 1$ MeV and $\Gamma\approx0.5$ MeV) 
dipole resonance \footnote{As already stated in Sect.\ref{sect:intro}, concerning the "continuum effect" introduced in ref. \cite{Catara:96,Nagarajan:05}, we refer to Sect. V}  has been observed carrying about $6-8\%$ of the TRK sum rule \cite{Sackett:93,Zinser:97,Kobayashi:89,Nakamura:06,Kanungo:15,Tanaka:17}. This mode, exchanged between the halo neutrons,
binds them to the core ($^9$Li), and provides the first example of a van der Waals nuclear Cooper pair (see App. A, Fig. A1(e) of \cite{Broglia:16}; see also Fig. 1(D)).

The size of folding domains of globular proteins ($\approx 125$ amino acids) is intimately connected to the dispersive (retarded) properties of van der Waals forces (cf. \cite{Israelachvili:10,London:30,Pauling:85} and refs. therein). Similarly, the exchange of pygmy resonance bosons between neutrons conditions the size of a nuclear halo Cooper pair. This bootstrap mechanism also provides a natural answer to the vexing question posed by the so-called pairing anti-halo effect (\cite{Bennaceur:00, Hamamoto:03, Hamamoto:04}.
In a nutshell, neutrons moving in orbits close to threshold with essentially no or little centrifugal barrier (i.e. $s_{1/2}$, $p_{1/2}$) have a large probability of being outside the core nucleus. Therefore, loosely bound neutrons become unavailable for pair correlation induced by the short range nucleon-nucleon, bare pairing force ($^{1}S_0$). This is because the matrix element $M_{\delta} = \langle j^2(0) | V_\delta | j^2(0)\rangle$ of a contact interaction is proportional to $1/R^3$ (see e.g. Sect. 2.3 of \cite{Brink:05}). Consequently, since $R\left(^{9}\textrm{Li}\right) \approx 2.5 \textrm{ fm}$, while $R\left(^{11}\textrm{Li}\right) \approx 4.58\pm0.13 \textrm{ fm}$, $M_\delta$ is reduced in the nucleus $^{11}$Li by a factor $(2.5/4.6)^3 \approx 0.2$ with respect to $^9$Li. 
In other words, one has a factor 5 reduction of the bare pairing interaction without an accompanying  change of the level density (see Sect. \ref{sec:microscopic}).

The poor overlap between core and halo neutrons makes the pairing interaction in $^{11}$Li subcritical, depriving the two--particle system of the necessary correlation energy. The fact that $^{11}$Li is bound, alas weakly ($\approx 380$ keV), testifies to the existence of a subtle, long wavelength pairing mechanism in which the corresponding intermediate boson is the soft $E1$--mode with centroid $\lesssim 1$ MeV. This is again a consequence of the poor overlap existing between the extended halo single-particle wavefunctions and those of the core, overlap which leads to a  screening of the repulsive symmetry potential, 
allowing for a conspicuous fraction of the dipole strength ($\approx 6-8\%$) to show up at low energies, almost degenerate with the ground state \cite{Barranco:01}.

Summing up, the highly extended low--momentum neutron halo of $^{11}$Li needs to vibrate against the core to give rise to a soft $E1$--mode. But to do so, the neutron--halo Cooper pair has to exchange this mode between its partner neutrons to get bound to $^9$Li. Neutron--halo pair addition mode and the soft $E1$--mode are thus, in $^{11}$Li two symbiotic elementary modes of excitation resulting from a bootstrap mechanism of incipient nuclear superfluidity.

In refs. \cite{Lenske:01} (see also \cite{Orrigo:09}), coupled-channel Gorkov techniques were used to describe the
pairing interaction between the two halo neutrons as well as core polarization in $^{11}$Li. The results provide an overall account of the experimental findings, in which the last two neutrons occupy $s$, $p$ and, to a small degree, also $d$-- states. If one transforms the pp representation used in \cite{Lenske:01} into the ph-channel, arguably one would obtain evidence for the quadrupole (self-energy) and dipole (induced pairing interaction) phenomena mentioned above and at the basis of \cite{Barranco:01}.
%physics at the basis of the results obtained resembles very much those of this reference.

One can then posit that the mechanism responsible for the presence of the  soft $E1$--mode in $^{11}$Li reflects an unusual embodiment of inhomogeneous damping phenomenon \cite{Bortignon:98,Dattagupta:87}. Inhomogeneous damping which, among other things, is at the basis of nuclear magnetic resonance (NMR) in general  \cite{Slichter:63}, 
and of studies of macromolecules in particular  (cf. e.g. \cite{Cavanagh:07,Rule:06,Lambert:04}, cf. also \cite{Roesner:17} and refs. therein), 
of surface plasmon resonance (SPR) research concerning surface enzyme kinetics \cite{Fang:05}, let alone rotational damping in nuclei \cite{Broglia:87}. 
However, and at variance with the above mentioned examples, in the case of  halo pygmy dipole resonance,
 the collective coordinate is the relative isotropic radial extension of the halo neutron field with respect to the nucleon saturated compact core. 
 In other words, in the present case, the inhomogeneity is not a function of angle but of the radial dependence of the 
 nuclear density (see Fig. 2 below) associated with the formation of a misty cloud around the core (neutron skin and neutron halo), 
 consistent with the lowering of the momentum of the last neutrons (cf. e.g. \cite{Hansen:93,Hansen:96,Austin:95}).

Let us elaborate in what follows on the nature of this soft dipole mode and of its consequences regarding the low-energy nuclear structure associated with incipient  - 
single Cooper pair -  nuclear superfluidity. \textit{In particular in connection with a novel interpretation of nuclear deformation} in halo nuclei, providing a new window to test the state dependence of the Axel-Brink hypothesis\footnote{Axel-Brink hypothesis: on top of each nuclear state one assumes there is built a GDR with the same properties as that built on the ground state. Most important in the study of the decay of highly excited nuclei (see e.g. Fig. 3 of \cite{Bertsch:86}) and thus also in connection with a central subject of today's research in nuclear physics: the level density.} \cite{Axel:62,Brink:PhD}

\section{Giant Dipole---, and Pygmy Dipole---Resonances}\label{sec:dipole}

The first vibrational mode to be observed in nuclei was the giant dipole resonance (GDR). It was excited by shining a beam of photons on a target (\cite{Baldwin:47,Baldwin:48}, see also \cite{Hirzel:47,Bothe:37,Bothe:39}). Even before then, it was recognized \cite{Migdal:44} that a mean excitation frequency for dipole absorption could be derived from the nuclear polarizability with the help of the symmetry energy in the mass formula \cite{Weizsacker:35}. Since then, this mainly isovector mode of nuclear excitation, in which protons and neutrons moving out of phase with respect to each other display an antenna-like oscillation with energy\footnote{This parametrization testifies the elastic character of the GDR mode. In fact the frequency of an elastic mode is given by, $\omega_{el}^2\sim\frac{\mu}{m\rho R^2}\sim\frac{v_t}{R^2}$, where $R$ is the radius of an elastic material sphere, $\rho$ the density and $v_t$ the transverse sound velocity proportional to the Lam\'e shear modulus of elasticity $\mu$, see p.43 of \cite{Bertsch:94}.} $\hbar \omega_{GDR} \approx 100 \textrm{ MeV}/(R_{0})_{fm}$, has been observed in essentially all nuclei throughout the mass table \cite{Bohr:75,Bortignon:98,Gaardhoje:92}.

In nuclei displaying neutron excess the admixture of isovector modes with isoscalar vibrations is unavoidable, as e.g. in the case of $^{208}_{82}$Pb$_{126}$ ($(N-Z)/A \approx 0.21$), let alone $^{11}_{3}$Li$_{8}$ ($(N-Z)/A \approx 0.45$). Such a mechanism studied since the seventies brings down part of the TRK sum rule, the resulting strength being called ``pygmy'' resonance \cite{Lane:71,Starfelt:64}. Within this context, the consequences of the admixture of isoscalar and isovector modes in the case of another giant vibration, namely the Giant Quadrupole Resonance (GQR)  were discussed in detail in \cite{Bes:75a}. 

During the last years a large \textit{corpus} of studies on low-energy $E1$-modes has been accumulated 
(see e.g. \cite{Ryezayeva:02,Paar:07,Lanza:09,Lanza:11,Lanza:11a,Piekarewicz:12,Savran:13,Scheck:13,Baran:13,Brenna:13,Inakura:13,Ozel:14,Ponomarev:14,Roca:15,
Loher:16,Arsenyev:17,Martorana:18,Savran:18,Sun:18} and refs. therein). In what follows we discuss some aspects of the physics at the basis of PDR, helped by the unique insight  provided by  the soft $E1$--mode of $^{11}$Li.

\subsection{Mechanism to create a pygmy  dipole resonance (PDR)}\label{subsec:mechanism}

The basic condition to fulfill in the quest to create a PDR  is that of finding particle-hole excitations of dipole character 
displaying a poor overlap with those building up the GDR. 
Arguably, the only way to do so is through $(p,h)$ 
excitations in which the particle moves in  an $s_{1/2}$ state essentially at threshold $(\approx S_n)$. In this way the associated component can partially tunnel out of the system and visit regions with a density radius somewhat larger than that experienced by the $(p,h)$ components of the GDR wavefunction. To avoid the Coulomb barrier,
 these excitations have to be of neutron type.
 % {\color{green} pero' c'è l'esempio di 100Sn} 
 Because the GDR has quantum numbers $1^-$, two are the possible $(p,h)$ excitations of this type: $(p^{-1}_{1/2},s_{1/2}), (p^{-1}_{3/2},s_{1/2})$. 
 Due to spin-orbit effects, if one is operative then the other will not.
Other $(p,h)$ configurations can, mixing with the $(p^{-1},s)$ components, 
explore regions of space of radius larger that $R_0 = 1.2 A^{1/3}$ fm, the so-called neutron skin region 
and, in this way lower their confinement (kinetic) energy. This is so provided the associated single-particle states experience moderate centrifugal barriers. A condition which restricts the choices essentially to the $(p^{-1},d)$ and $(f^{-1},d)$ components.
Examples of nuclei displaying a sizeable  fraction of the TRK sum rule at low energy are $^{208}$Pb and $^{122}$Sn.

In $^{208}$Pb the PDR has been observed not far from threshold ($S_n = 7.37$ MeV), its energy being $\approx 6$ MeV, similar to $\hbar \omega_0 \approx 6.9$ MeV, the energy difference between major shells. The main components of this mode are calculated to be $(3p^{-1},4s)$, $(3p^{-1},3d)$, and $(2f^{-1},3d)$ \cite{Ryezayeva:02,Lane:71}. In the case of the isotopes $^{112}$Sn--$^{130}$Sn, the wavefunctions of the lowest dipole states of a systematic QRPA calculation ($E_x = 7.9 - 5.8$ MeV) carrying a consistent $B(E1)$ strength and thus being part of the PDR, have as main component the neutron configuration ($3s_{1/2}$, $3p_{3/2}$) \cite{Tsoneva:08}.  In $^{11}$Li, the soft dipole mode has been observed at a very low energy $\lesssim 1$ MeV, close but above threshold ($S_{2n} = 369$ keV), the main components being of type $(p^{-1},s)$.

Within this context one can make reference to \cite{Ryezayeva:02} (see also \cite{Roca:15,Savran:13,Paar:07}), where a detailed experimental and theoretical analysis of high resolution $\gamma,\gamma'$ study of the electric dipole response in $^{208}$Pb is reported. A resonance-like structure was observed close to the neutron threshold which is interpreted as the PDR. Microscopic quasiparticle-phonon model  \cite{Ryezayeva:02} calculations (\cite{Paar:05,Tamii:11} see also \cite{Bortignon:81} and \cite{Bertsch:83}) which provide an overall account of the experimental data, suggest that this mode 
can be viewed as an oscillation of the neutron skin against an approximately isospin-saturated core. The associated velocity distribution is dominated by a vortex-like pattern.
%{\color{green} Pero' Reinhard trova che questo campo di velocita' e' incompatibile con l'oscillazione core-halo} .The similitude of these finding with the results of our study of the soft $E1$--mode of $^{11}$Li \sout{are} {\color{red} is}  taken up in detail in Sect. V.

Let us conclude this section by noting that, if one would like to transform essentially any stable nucleus into systems displaying low--energy dipole modes, one has just to warm the system up, and study the $\gamma$--decay of the corresponding compound nucleus. Recent theoretical results \cite{Wibowo:18,Litvinova:18} indicate the presence of PDR in hot nuclei at $T \approx 3$ MeV, e.g. in $^{68}$Ni, are also described by wavefunctions made out of $(3s_{1/2}$, $3p_{3/2}$) and ($2d_{5/2}$, $3p_{3/2}$) neutron components (for more detail see Sect. \ref{sec:dipole}).

\subsection{Mean Field Models}\label{subsec:mean}
%{\color{green} Non capisco il titolo di questa sezione }

Because of the large transition moments associated with the vibrational excitations and as a result of  the particle-vibration coupling, the single-particle 
states become clothed moving   in a cloud of quanta (\cite{Bohr:75} p.420). What we call a nucleon, can be physically described part of the time by the action of the bare nucleon,
but another fraction of the total history also involves the action of the collective vibrations. Within this framework, the distinction  between single-particle 
and collective motion  is set in the proper   perspective  without indulging in the pervasive and according to \cite{Mottelson:62}  perverse sharp antithetical  connotation found in the literature,
to emerge as two deeply interweaved complementary aspects  of the finite many-body nuclear problem. This is the reason which justifies to treat the E1-decay  of the  PDR
of $^{11}$Li and that connecting the $\widetilde{1/2}^- \to \widetilde{1/2}^+$ transition of $^{11}$Be, and parity inversion found at the basis of them, on equal footing. 

As mentioned  previously, the distinction between elastic- and plastic-like collectivity has profound physical consequences in the low-energy nuclear
structure. For example the coupling to the GDR is of little importance concerning the single-particle self energy, while the polarisation contribution to the effective 
charge can lead to almost an order  of magnitude reduction of it. In the process, the GDR builds up its EWSR content. Essentially the only way a single-particle 
transition can avoid losing its "collective" $\approx$ 1 $B_W (E1)$ character, is to be associated  with single-particle wavefunctions displaying  a small overlap 
with those which build up the GDR. That is, the scenario of neutron skin in general and of the neutron halo in particular.  

Concerning a plastic mode, e.g. a low-lying quadrupole vibration,  displaying an energy $\hbar \omega $  much smaller than the distance between single-particle states, the self energy effect  on the single-particle  can be conspicuous and approach the static limit.
Within this context one can mention parity  inversion in $^{11}$Be and make reference to \cite{Shimoura:03,Shimoura:07,Hamamoto:07,Macchiavelli:18}.
However, when the frequency of the plastic mode, like for example the quadrupole vibration of the core $^{10}$Be, is of the order of the 
distance between the single-particle levels, the $\omega-$dependence of the renormalization processes has to be taken explicitly into account. Let alone 
the induced interaction in the case of a two-particle system arising from the exchange of the quadrupole vibration ($^{12}$Be), or of the  soft $E1$--mode($^{11}$Li). 
Especially,  because in such a situation, the system can be close to a resonant phenomenon. {\it Resonant behaviour which is expected to be found in e.g. the absolute differential cross sections associated with the reactions $^{11}$Be(p,d)$^{10}$Be($2^+$) and $^9$Li(t,p)$^{11}$Li(gs) as a function of  the proton and triton bombarding energies respectively}.  Phenomenon observed in electron tunnelling across a junction between metallic superconductors as a function of the 
biasing potential, and providing information on the electron-phonon coupling (inversion of Eliashberg equations \cite{Schrieffer:64}). Let us now return 
to the discussion of the static limit in the case of $^{11}$Be.

It could be argued that in  $^{11}$Li the pygmy resonance carrying of the order of one Weisskopf unit can be viewed as a single-particle $s \rightarrow p$ transition 
as, for example, in the case of the $\widetilde{1/2^-} \to \widetilde{1/2^+}$ transition in $^{11}$Be discussed in \cite{Hamamoto:07,Macchiavelli:18} (see also \cite{Shimoura:03,Shimoura:07}). 
Parity inversion results, in the model used,  from a static, axially symmetric quadrupole deformation. This makes the $1/2^+ ([220 1/2])$ downsloping component of the $1d_{5/2}$ orbital and the upsloping $1/2^- ([101 1/2])$ $1p_{1/2}$ state to invert parity at $\beta_2 \approx 0.6$ (Nilsson levels cf. Fig 5.1 p. 221 of ref. \cite{Bohr:75}).
This is not surprising, as collective and single-particle motion emerge from the same features of the nuclear forces  \cite{Mottelson:62}.
%In a sense, independent particle can be considered the most collective feature of nuclear structure as it requires the coherent participations 
%of all nucleons in generating the mean field

In spite of its attractiveness, and to the extent the validity of the parallel made between E1-transitions in $^{11}$Li and in $^{11}$Be
holds, 
such a picture seems to be contradicted by experiment. In fact, the dipole resonance populated in the $^{11}\textrm{Li}(d,d')^{11}$Li($1^-$) 
reaction, with centroid $E_X = 1.03 \pm 0.03 \textrm{ MeV}$ and width (FWHM) of $0.51 \pm 0.11 \textrm{ MeV}$ \cite{Kanungo:15} displays a single peak. This testifies against the presence of a static quadrupole deformation (see \cite{Bohr:75} Fig. 6-21, p. 491). Furthermore, such a description of $^{11}$Be  
lacks some important physics: the exclusion principle, and the tendency -dynamical instability- of the system to acquire a permanent dipole moment.
To come down to threshold, the $2s_{1/2}$ couples strongly to the quadrupole vibration of the core $^{10}$Be, mainly through the $1d_{5/2}$ state, leading to a self-energy (polarization) energy shift of $\approx -1$ MeV. 
Because the main component of this quadrupole vibration corresponds to the ($1p^{-1}_{3/2}$--$1p_{1/2}$) configuration ($X_{1p_{3/2}-1p_{1/2}} = 1.02$ \cite{Barranco:01}), the self energy (correlation) process resulting from the exchange of the $1p_{1/2}$ nucleon and the same nucleon participating in the vibration 
leads to a repulsive, exclusion principle, correction and to an upshift in energy of $\approx +2$ MeV. In the process, the $|\widetilde{1/2}^-\rangle$ state becomes
almost  degenerate with the 
$|\widetilde{1/2}^+\rangle$ state, although higher in energy by about 300 keV (parity inversion).
As a result of the dressing process, the single-particle content of both the ${1/2}^+$ and ${1/2}^-$ states is only $\approx 80\%$, as needed to reproduce the absolute differential cross sections observed in $^{10}$B(d,p) and $^{11}$Be(p,d) (see
\cite{Winfield:01,Barranco:17} and refs. therein.)
%Within the above picture, and at variance with \cite{Hamamoto:07,Macchiavelli:18} and similar approaches which require the use of two single-particle potentials, one for positive parity and another for negative parity orbitals, a single potential is used in the nuclear field theory (NFT) description of $^{11}$Li \cite{Barranco:01}. 

In connection with the single-particle states  at the basis of the neutron degrees of freedom of $^{11}$Li, i.e. $^{10}$Li, 
the $|\widetilde{1/2^+}>$  and $|\widetilde{1/2^-}>$ 
renormalised states provide an overall account of the single-particle content observed in $^9$Li$(d,p)^{10}$Li and $^{11}$Li$(p,d)^{10}$Li reactions (see \cite{Jeppesen:06,Orrigo:09,Sanetullaev:16,Casal:17}; see also \cite{Cavallaro:17} and \cite{Barranco:tbp}).

% \begin{table}[h!]
% \begin{center}
% \begin{adjustbox}{max width=\textwidth}
% \begin{tabular}{|c|c| c |  c | c |  c |  }
% \hline 
% & $n_1l_1j_1$ &$n_2l_2j_2$& $X$ 
% %& $a_{WS}$  &  $R_{WS}$   \\ \hline
% %$^{11}$Be& 0.78 & 68.9 & 0.47 & 0.77& 2.15 \\ \hline
% %$^{12}$B&  0.59 & 74.2&  0.60 & 0.76 &2.20 \\ \hline
%  %$^ {13}$C  & 0.46  & 79.5 & 0.7& 0.75  & 2.25    \\ \hline
%  \end{tabular}
% \end{adjustbox}
% \end{center}
% \caption{   QRPA wave functions of states representative of the GDR and of the PDR. In the first column
% the quantum numbers of the two quasiparticles are given ... } 
%  \label{Table1}
% \end{table}
% 

Concerning the dipole instability and associated conspicuous dipole zero-point fluctuations, the phenomenon is common to nuclei around the $N=6$ closed shell system (see e.g. \cite{Potel:14} and refs. therein). In particular one can mention \footnote{See \cite{Barranco:17,Liu:12,Calci:16} for an analysis of the interplay between $E1$ transitions, parity inversion and continuum in Be isotopes.}
: a) $^{10}$Be in which case the lowest $1^-$ state is almost degenerate with the first (second) $0^+ (2^+)$ states at $\approx 6$ MeV, and thus essentially at threshold ($S_n = 6.81$ MeV); b) $^{11}$Be, in which case $|\widetilde{1/2}^-\rangle$ state is found at 0.32 MeV above the  $|\widetilde{1/2}^+\rangle$ ground state, with $S_n = 501$ keV and a dipole $B(E1)$-value 
 close  to one Weisskopf unit ($B_{sp}(E1)$; c) $^{12}$Be, in which case the first excited $0^{+}$ (2.24 MeV) and first excited $1^-$ state (2.70 MeV), likely a fragment of the soft  dipole mode, are rather close in energy, the $1^-$ state being not far from threshold ($S_n = 3.17$ MeV); d) $^{11}$Li in which case one observes a concentration of dipole strength with centroid $\lesssim 1$ MeV and a width of about 0.5 MeV ($S_{2n} \approx 369$ keV \cite{Kanungo:15}), the associated inelastic cross section not being inconsistent with a $B(E1)$ of the order of one single-particle unit.

The above examples, provide a scenario which resembles that described by Lane \cite{Lane:71} when confronted with the unexpected positive correlation of neutron and partial photon widths of neutron resonances. Quoting: ``the resolution of this problem arises from the existence of the ''giant'' and "pygmy" dipole resonance ... The vital question is: what is the nature of the small amount of $E1$ strength in the threshold region? There are two sources of this: (i) the effect of random dissipative forces on the collective state broadens it, and introduces a random "tail"  into the threshold region,
(ii) a systematic residuum of $1p-1h$
strength which belongs to the threshold region. The data on $(n, \gamma)$ spectra at $185 < A < 208$ shows an anomalous bump which has been interpreted \cite{Starfelt:64} as a "pygmy" dipole resonance centred at $E_\gamma \approx 5.5$ MeV. 
This shows that (ii) ... can be regarded as arising from a single
collective state which is the doorway state for photons in the threshold region. 
The only remaining question is whether this state has a large component of the $s\rightarrow p$
transition ... a calculation of $^{208}$Pb strongly suggests that it is large. ''

It is important to mention that the QRPA wavefunction of the $E1$--soft dipole resonance ($E_x \approx 0.75$ MeV) of $^{11}$Li, aside from having a large $(1p^{-1}_{1/2},2s_{1/2})_{1^-}$ component ($X_{1p^{-1}_{1/2}-2s_{1/2}} \approx 0.847$, see Fig. \ref{fig:3} below as well as Table 1), has a number of other components implying p--h jumps across major shells, typical of the $\Delta N = 1$ component of the GDR.
In fact, the value of these components as well as their relative phases play a central role in screening the  pygmy dipole resonance from being depleted from $E1$-strength by the GDR, allowing this mode to retain a consistent fraction of the TRK sum rule ($\approx 6-8\%$).

One can then posit that without a screening factor (small overlap), most of the $(s \rightarrow p)$-$E1$ strength will have been shifted to higher energies (repulsive character of the symmetry potential). In fact, typical $E1$ transitions between pure single-particle orbitals of nuclei lying along the stability valley are of the order of $\lesssim 10^{-2}$ Weisskopf units.
 
Summing up, it can be stated that the above microscopic description of the $^{11}$Li soft dipole mode provides a physical picture of the mechanism described in point (ii) of ref. \cite{Lane:71}. Furthermore, it is well established that the damping width of giant resonances, including also an eventual  long tail, 
arises  from the coupling to doorway states made out of an uncorrelated particle-hole excitation and a collective low-lying vibration \cite{Bortignon:81,Bertsch:83} 
%(quasiparticle-phonon model \cite{Paar:05,Tamii:11}). 
(see also \cite{Ryezayeva:02}).
Damping which, for collective states like the GDR, is strongly suppressed due to cancellations. Consequently, mechanism (i) can hardly be the origin of a soft dipole mode. On the other hand, this cancellation and resulting reduced width provides, arguably, a solution to the (quoting again from \cite{Lane:71}):``... striking inconsistency between the calculations and recent data on the $E1$ widths of $1^-$ states of $^{208}$Pb in the region near 5.5 MeV.''
This is in keeping with the fact that the calculations referred to were  based essentially on uncorrelated damping 
of particles and holes, thus predicting widths much larger than experimentally observed.

Aside from the above issues we are interested, in the present paper, to shed light on a somewhat ignored, but central role played by the soft $E1$--mode in the low-energy nuclear structure spectrum, essentially reflecting its large amplitude, plastic properties (cf. \cite{Brink:05,Barranco:01,Broglia:01,Broglia:10}).
In other words, in the interweaving of dipole modes essentially degenerate with the ground state and single-particle motion, in particular as intermediate boson. \textit{As a bonus, one acquires what can be considered the specific probe of an important type of nuclear deformation: isotropic radial deformation. Such deformation is closely connected with the neutron skin leading to incipient and/or well defined halo structures}.

\section{Elementary Modes of Nuclear Excitation}
\label{sec:elementary}

Nuclear structure can be described in terms of elementary modes of excitation (cf. \cite{Bohr:75, Mottelson:76, Bes:77} and refs. therein). Each elementary mode is associated with a different ground state (Fig. \ref{fig:1}). Hartree-Fock ($a_k |HF \rangle = 0$, $[H,a^\dagger_i] = \varepsilon_i a^\dagger_i$; $\varepsilon_i < \varepsilon_F$, $\varepsilon_k > \varepsilon_F$) for single particle motion, i.e. elementary modes of excitation obtained adding or removing a particle to or from a nuclear orbital, and carrying transfer quantum numbers $\beta = \pm 1$. RPA ($\Gamma_\alpha (\beta = 0) |\tilde 0 \rangle_v = 0$, $\Gamma^\dagger_\alpha (\beta = 0) = \sum_{k i} (X_{k i} \Gamma^\dagger_{k i} + Y_{k i}\Gamma_{k i})$, $\Gamma^\dagger_{k i} = a^\dagger_k a_i$, $[H,\Gamma^\dagger_\alpha] = \hbar \omega_\alpha \Gamma^\dagger_\alpha$) 
for collective particle-hole-like vibrations, as well as for pairing vibrations ($\Gamma_\alpha (\beta = \pm 2) |\tilde 0 \rangle_{pv} = 0$, $\Gamma^\dagger_\alpha (\beta = +2) = \sum_{k} X_{k} \Gamma^\dagger_{k} + \sum_{i} Y_{i}\Gamma_{i}$, where $\Gamma^\dagger_{k} = a^\dagger_k a^\dagger_{\tilde k}$ and $\Gamma_{i} = a^\dagger_i a^\dagger_{\tilde i}$, 
and similarly for $\Gamma^\dagger_\alpha (\beta = -2)$). Thus, the nuclear ground state 
can, within the subspace of these modes (also called collective coordinates in other fields of many-body research) and to a good approximation, be written as
\[
|\tilde 0 \rangle = | HF \rangle \otimes |\tilde 0 \rangle_v \otimes |\tilde 0 \rangle_{pv},
\]
in the case of nuclei displaying double or single closed shells and as (intrinsic system),
\[
|\tilde 0 \rangle = | \textrm{Nilsson} (HFB) \rangle \otimes |\tilde 0 \rangle_{\beta, pv} \otimes |\tilde 0 \rangle_{\gamma},
\]
in the case of quadrupole deformed superfluid open shell nuclei displaying $\beta$-- and pairing-vibrations ($K=0$) which mix, and $\gamma$-vibrations ($K=1$) \cite{Bohr:75}. Elementary modes of excitation contain a large fraction of the nuclear correlations, their interweaving being amenable to a field theoretical treatment (see \cite{Mottelson:76,Bes:77,Bortignon:77,Bes:75,Bortignon:78} and refs. therein).
Within this scenario one can posit that in order that two different \textit{bona fide} elementary modes of excitation with the same quantum numbers and thus ground state, like e.g. the giant dipole and the pigmy  dipole resonances (cf. \cite{Savran:13,Scheck:13,Baran:13,Piekarewicz:12} and refers. therein) can coexist is that they are essentially build out of single-particle wavefunctions and of particle-hole excitations which display a small overlap $\mathcal{O} \ll 1$).

\section{Microscopic description of the soft $E1$--mode of $^{11}$Li}\label{sec:microscopic}

Let us now work out the microscopic wavefunction of the soft $E1$--mode of $^{11}$Li, within the framework of QRPA. 
To be able to do so we have to  to calculate the $U,V$ occupation factors which, in the present case,
as we explain below, is a rather subtle requirement.

Observations indicate that the mean square radius  of $^{11}$Li is $\left< r^2 \right>^{1/2} = 3.55 \pm 0.1$ fm \cite{Kobayashi:89}. Thus:
\begin{equation}
 R \left( ^{11} \textrm{Li} \right) = \sqrt{5/3} \left< r^2 \right>^{1/2} = 4.58 \pm 0.13 \textrm{ fm}.
 \label{eq:RLi}
\end{equation}

Making use of the radius expression $R_0 = 1.2A^{1/3}$ fm obtained from systematics of nuclei along the stability valley leads to $R_0 = 2.7$ fm ($A=11$), while the value $R = 4.58$ fm corresponds to an effective mass number $ \approx 56$, five times larger than the actual value.

\begin{figure}[ht!]
	\begin{center}
		\includegraphics[width=0.85\textwidth]{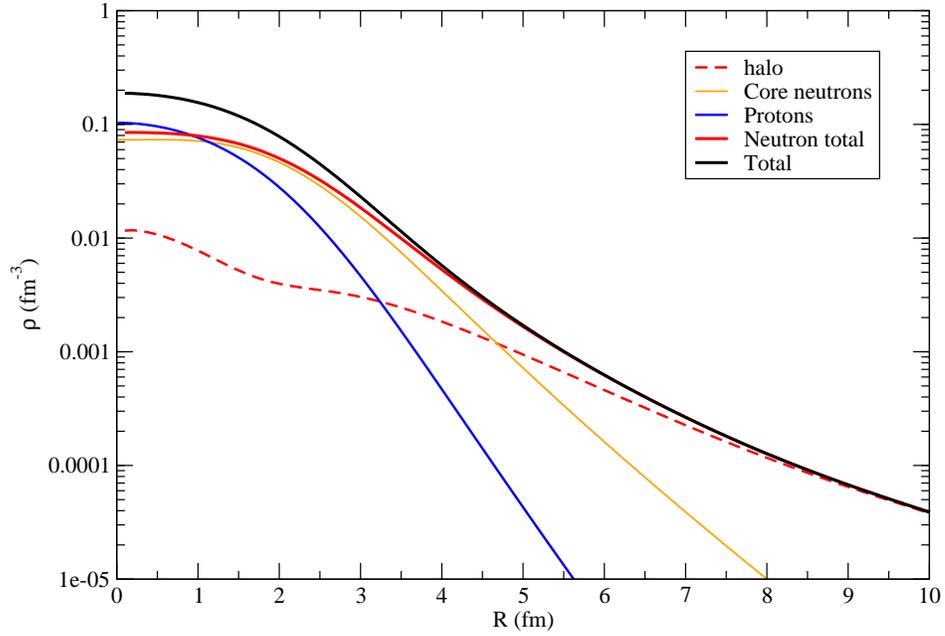}
	\end{center}
	\caption{Density of $^{11}$Li calculated within the framework of nuclear field theory for the renormalisation of the single-particle levels close to threshold (see Fig. 1(D)). The halo refers to the wavefunction (5) (modulus squared). See also Fig. 3. }
\label{fig:2}
\end{figure}

\begin{figure}[ht!]
	\begin{center}
		\includegraphics[width=0.75\textwidth]{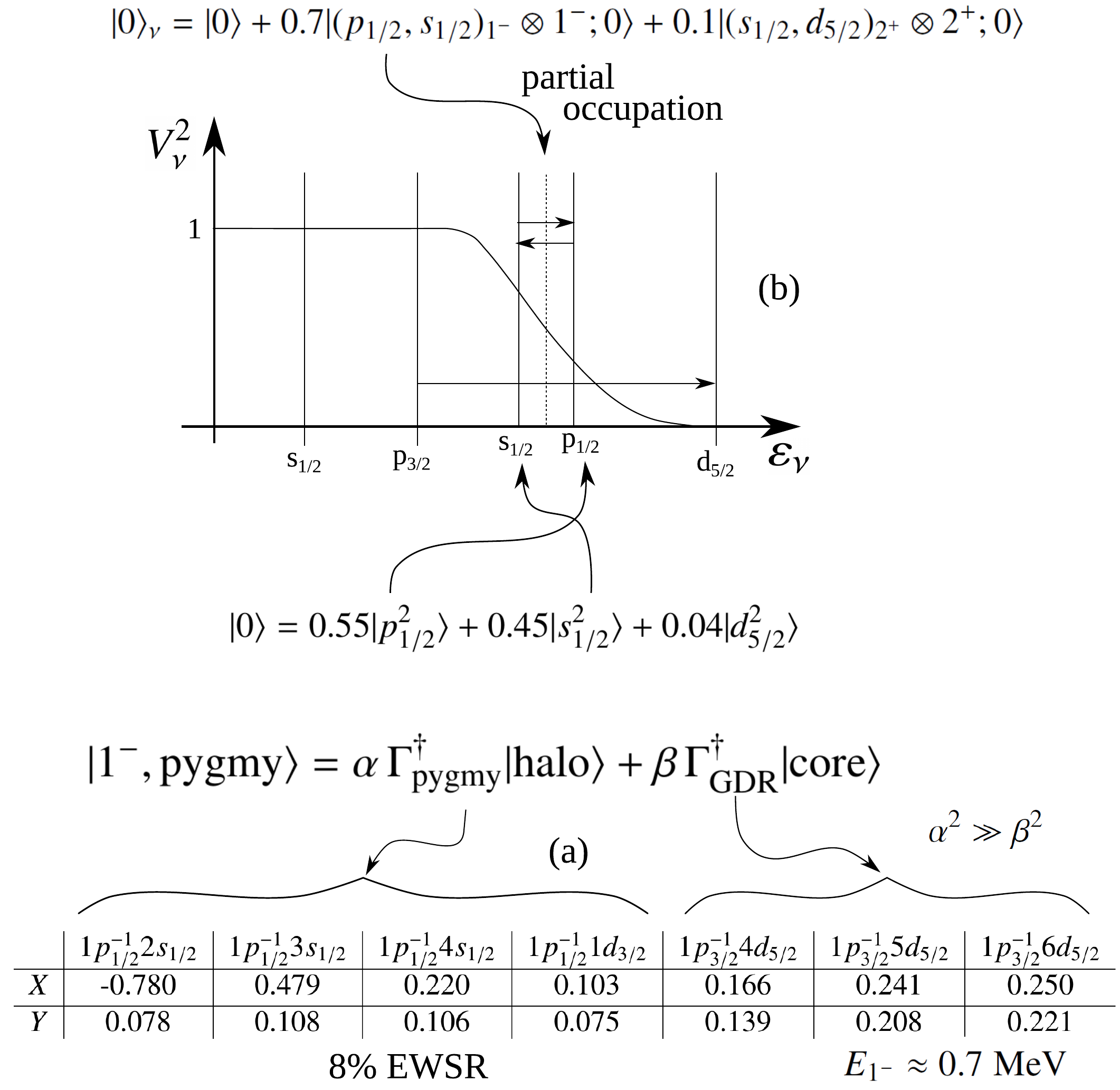}
	\end{center}
	\caption{Schematic representation of (a) the QRPA calculations of the soft dipole mode (labeled pygmy for convenience) of $^{11}$Li and associated results. Namely $X$ and $Y$ QRPA amplitudes divided, for didactic purposes, into low-lying (pygmy) and high lying (GDR) $p-h$ excitations. It is of notice that throughout the $p_{3/2}$ proton state is not shown being treated as a spectator; (b) schematic representation of the connection between occupation numbers and nuclear field theory wavefunction describing the two halo neutrons. The wavefunction of the state is expressed in the laboratory system, immaterial as far as the $E1$--distribution strength is concerned, the $B(E1)$ strength being calculated making use of a dipole operator which eliminates the contributions of the center of mass motion.	}     
\label{fig:3}
\end{figure}

The above result testifies  to the very large isotropic radial deformation of $^{11}$Li. Let us parametrize the radius of $^{11}$Li as $R = R_0 \left( 1+\beta_0Y_{00} \right)$ \cite{Bohr:75} (see Eq. (3.14) of \cite{Bortignon:98}).
Thus $\beta_0 = \sqrt{4 \pi}\left( \frac{R}{R_0}-1\right) \approx 2.5$, which testifies to the extreme exoticity of the phenomenon, associated with a neutron skin (halo) thickness of the order of  $ R(^{11} \textrm{Li}) - R_0(^{9}\textrm{Li}) = R_0(^{9}\textrm{Li}) \left( \frac{R(^{11} \textrm{Li})}{R_0(^{9}\textrm{Li})}-1\right) \approx 0.8 R_0(^{9}\textrm{Li})$, where $R_0(^{9}\textrm{Li}) = 2.5$ fm is the radius of the $N=6$ closed shell core of $^{9}$Li.
In other words, $^{11}$Li can be viewed as built out of a normal $^{9}$Li core and of a skin made out of two neutrons moving around the core in a spherical shell of a range of the order 80\% the core radius. 
As a result one is dealing with a rarefied neutron atmosphere of density, 
\begin{equation}
\rho \approx \frac{2}{ \frac{4 \pi}{3}\left( R \left( ^{11} \textrm{Li} \right) \right)^3 - \frac{4 \pi}{3} \left( R_0 \left( ^{9} \textrm{Li} \right) \right)^3} \approx 0.6 \times 10^{-2} \textrm{ fm}^{-3},
 \label{eq:dens}
\end{equation}
that is, approximately 4\% of saturation density . Within this context $\langle r^2 \rangle_n = 14.5$ fm$^2$ and $\langle r^2 \rangle_p = 4.4$ fm$^2$, leading to a neutron skin of value $\Delta r_{np} \approx 1.71$ fm while $\langle r^2 \rangle_{\textrm{total}} \approx 11.8$ fm$^2$ (see Fig. \ref{fig:2}).

Because $^{10}_{3}\textrm{Li}_{7}$, with one neutron outside closed shell is not bound, while adding a second neutron results in the bound $^{11}$Li system, one is confronted with a neutron pairing phenomenon at very low nuclear density. The fact that the two neutron separation energy of $^{11}$Li is $S_{2n}=380$ keV , as compared with $\approx 15$ MeV for stable nuclei, as well as the large dimensions of the halo dineutron system, testifies to the fact that we are in presence of a nuclear embodiment of Cooper pair model \cite{Cooper:56}.

%{\color{green} Secondo me questa parte va riscritta. A p. 7 l'argomento e' scritto in modo piu' semplice e porta subito a uno screening di 0.2. Il fattore 1/4 dovuto a j  che entra qui non lo capisco, 
%lo screening va fatto rispetto ad un'interazione in un nucleo di raggio standard ma sempre in s-waves. Inoltre $V_0$ non e' definito e comunque non e' uguale a quello del libro con Brink.

In fact, the two neutrons move in the continuum, in particular in the virtual 
\footnote{$|\epsilon|= \frac{\hbar^2 \kappa^2}{2m}$ where $\kappa = -1/\alpha$, $\alpha = - \textrm{lim}_{k \to 0} \textrm{tg}(\delta_0)/k$ being the scattering length (\cite{Landau:65} p.507).}
$\widetilde{1/2^+}$ ($\approx 0.3$ MeV) and resonant $\widetilde{1/2^-}$ ($\approx 0.5$ MeV) states \cite{Barranco:tbp}, on top of a Fermi sea ($^{9}$Li core). 
Let us now switch on the pairing interaction schematically represented by a contact interaction. 
The matrix elements $\langle j^2(0) | V_{\delta} | j^2(0) \rangle = \frac{2j+1}{2}G$, where $G = 3 V_0/R_0^3 \approx 28/A$ MeV, and $R_0 = 1.2 A^{1/3}$ fm, while $V_0 = 
\frac{294}{4 \pi}$ MeV fm$^{-3}$ (see  Sect. 2.3 of \cite{Brink:05}). 
Consequently, in the case of $^{11}$Li, the strength will be screened out by a factor \footnote{The estimate of $2j+1$ for a ``normal'' system can be carried out making use of $k_F R\approx 1.36\textrm{ fm}^{-1} \times 2.7\textrm{ fm}\approx 3.7   $ and thus $\left(2\times3.7+1\right)\approx  8 $.} $\frac{2}{2j+1} \left(  \frac{R_0 \left( ^{11}\textrm{Li} \right)}{R \left( ^{11}\textrm{Li}\right)}\right)^3\approx \left(1/4\right) \times 0.2 \approx 0.05$. 
In other words, $G\left( ^{11}\textrm{Li}\right)=0.05\left(28/11\right) \textrm{ MeV} \approx 0.05 \times 2.5 \textrm{ MeV} \approx 0.13$ MeV. 
This strength of the pairing interaction is subcritical, the associated matrix element $\langle s_{1/2}^2 (0) | V_{\delta} | s_{1/2}^2 (0) \rangle = -130$ keV, not being able to bind the lowest configuration $|s_{1/2}^2 (0) \rangle$ of energy $2\varepsilon_{s_{1/2}} \approx 0.6$ MeV.
In fact, binding takes place essentially through the induced pairing interaction, the intermediate boson being the the soft dipole mode which exchanged between the two halo neutrons, binds them to the core $^{9}$Li (see Fig. \ref{fig:1} (D)).

Now, to work out the corresponding induced pairing matrix elements, one needs to have a microscopic description of the PDR. 
But to calculate it one needs to know what the occupation numbers of the single-particle states around the Fermi energy are, that is, to have a microscopic description of the $| gs \left( ^{11}\mathrm{Li} \right)\rangle$ state (bootstrap process). More precisely, writing
\begin{equation}
| gs \left( ^{11}_{3}\textrm{Li}_8 \right) \rangle = | \widetilde{0} \rangle_{\nu}\otimes | p_{3/2} (\pi) \rangle,
 \label{eq:gs}
\end{equation}
where $|p_{3/2}(\pi)\rangle$ is the odd proton considered to act as a spectator,
\begin{equation}
 |\widetilde{0}\rangle_{\nu} = |0\rangle + \alpha|(p_{1/2},s_{1/2})_{1^-} \otimes 1^-;0\rangle+\beta|(s_{1/2},d_{5/2})_{2^+}\otimes2^+;0^+\rangle,
\end{equation}
and 
\begin{equation}
| 0 \rangle = a|p_{1/2}^2 (0)\rangle + b|s_{1/2}^2 (0)\rangle +c|d_{5/2}^2(0)\rangle,
 \label{eq:onu}
\end{equation}
is the wavefunction of the two halo neutrons moving around the $N=6$ closed shell core $^{9}_{3}\textrm{Li}_6$, $a^2$, $b^2$ and $c^2$ providing the occupation probabilities needed to carry out a quasiparticle RPA (QRPA) calculation of the soft dipole mode.

The first steps in the self consistent process can be started with a BCS-like occupation distribution, calculate the amplitude $a$, $b$, $c$ and repeat the process now using the square of these amplitudes, until convergence is achieved. 
The outcome of these calculations regarding the neutron states of  $^{10}$Li is: 
\begin{equation}
 |\widetilde{1/2}^{+}\rangle=\sqrt{0.98}|s_{1/2}\rangle + \sqrt{0.02}|\left( d_{5/2} \otimes 2^{+}\right) {1/2}^+\rangle,
 \label{eq:12+}
\end{equation}
\begin{equation}
 |\widetilde{1/2}^{-} \rangle = \sqrt{0.94} | p_{1/2} \rangle + \sqrt{0.06} | \left( ( p_{1/2}, p_{3/2}^{-1} )_{2^+}\otimes 2^+ \right)_{0^+} , p_{1/2} ; {1/2}^- \rangle,
 \label{eq:12-}
\end{equation}
%\begin{equation}
%    \begin{split}
  %  |\widetilde{5/2}^+\rangle= {} &\sqrt{0.60}|d_{5/2}\rangle+\sqrt{0.20}|(s_{1/2}\otimes2^+){5/2}^+\rangle +\sqrt{0.15}|(d_{5/2}\otimes 2^+){5/2}^+\rangle+\\&+\sqrt{0.05}|(d_{3/2}\otimes2^+){5/2}^+\rangle,
  %  \end{split}
%\label{eq:52+}
%\end{equation}
where $|2^+\rangle$ stands for the low-lying quadrupole collective vibration of $^{9}Li$ ($\hbar\omega_2\approx3.3$ MeV, $\beta_2\approx0.7$).

In practice, a simplified description was carried out, making use of Saxon-Woods potentials to approximately reproduce both neutron and 
proton single-particle states, and BCS to determine the associated occupation numbers (see ref. [1]). The continuum was discretized 
%The calculations were carried out {\color{red} as}   discussed in ref. \cite{Barranco:01}, by including in the case of (\ref{eq:gs}) all the two--particle configurations $|nlj \times n'lj (0) \rangle$ with energy up to {150 MeV}, and with $n'$ different or equal to $n$, discretizing the continuous
 by placing the system in a spherical box of radius $R_{box} = 40$fm. Convergence was controlled by changing the value of $R_{box}$. It is of notice that Lenske et al. \cite{Lenske:01} found that the method of discretization of the continuum leads to results which reproduce those obtained through the exact treatment of the continuum, provided $R_{box}$ is about ten times the nuclear radius of $^{11}$Li ($\approx 4.6$ fm). The QRPA solution of the full dipole response calculated using a separable dipole-dipole interaction of self consistent strength was carried out in a two--quasiparticle basis with energies again up to 50 MeV \cite{Barranco:01,Bohr:75}.

A summary of the results associated with $^{11}$Li are displayed, for both the $(gs)$ and the soft $E1$--mode, in Fig. \ref{fig:3}. 
The interplay  between (dynamical) quadrupole deformations and the soft $E1$--mode  plays, in this barely bound nucleus, a central role, as observed by the $^{10}$Li single--particle states (Fig. \ref{fig:1}(C)) and by the $^{11}$Li ground state. Within this last connection, 50\% of this state corresponds to the component associated with the exchange of the soft dipole mode between the halo neutrons. Concerning the single-particle basis states, (6) and (7), one observes that between 2\%-6\% corresponds to configurations involving the $^9$Li quadrupole phonon, coupling responsible for parity inversion. While one would deem such probabilities quite small, one has to remember that
we are referring to continuum states. To compare with a similar set of states of the halo, bound nucleus $^{11}$Be, we have to take care 
of the  poor overlap ($ \mathcal{O}  \approx$ 0.2) existing between the virtual and resonant $|\widetilde{1/2^+} > $ and  $|\widetilde{1/2^-} > $ states of 
$^{10}$Li and those of the core $^9$Li. Thus, $0.02/ \mathcal{O}$   $\approx 0.1$ and 0.06/ $ \mathcal{O} \approx 0.3$  are the numbers which better compare to the
$^{11}$Be situation.

Finally in Tables I and II, the QRPA 
 wavefunctions  representative peaks of the soft dipole mode and of GDR are displayed.

\clearpage

\subsection{E1-strength function and transition densities}\label{subsec:e1}
Before  analyzing  the QRPA results of the full dipole linear response of $^{11}$Li, in particular the soft dipole mode, let us discuss the so--called ``continuum threshold effect'' \cite{Catara:96} (see also e.g. \cite{Paar:07} pp 717,718). 

The model used in \cite{Catara:96} considers the motion of a nucleon of mass $m$ in a spherical square--well potential, in which the $1p-1h$ transition strength can be calculated analytically. While the results presented mainly dwell with the unperturbed quadrupole response, unperturbed dipole and octupole responses are also considered.

\begin{figure}[ht!]
	\begin{center}
		\includegraphics[width=0.75\textwidth]{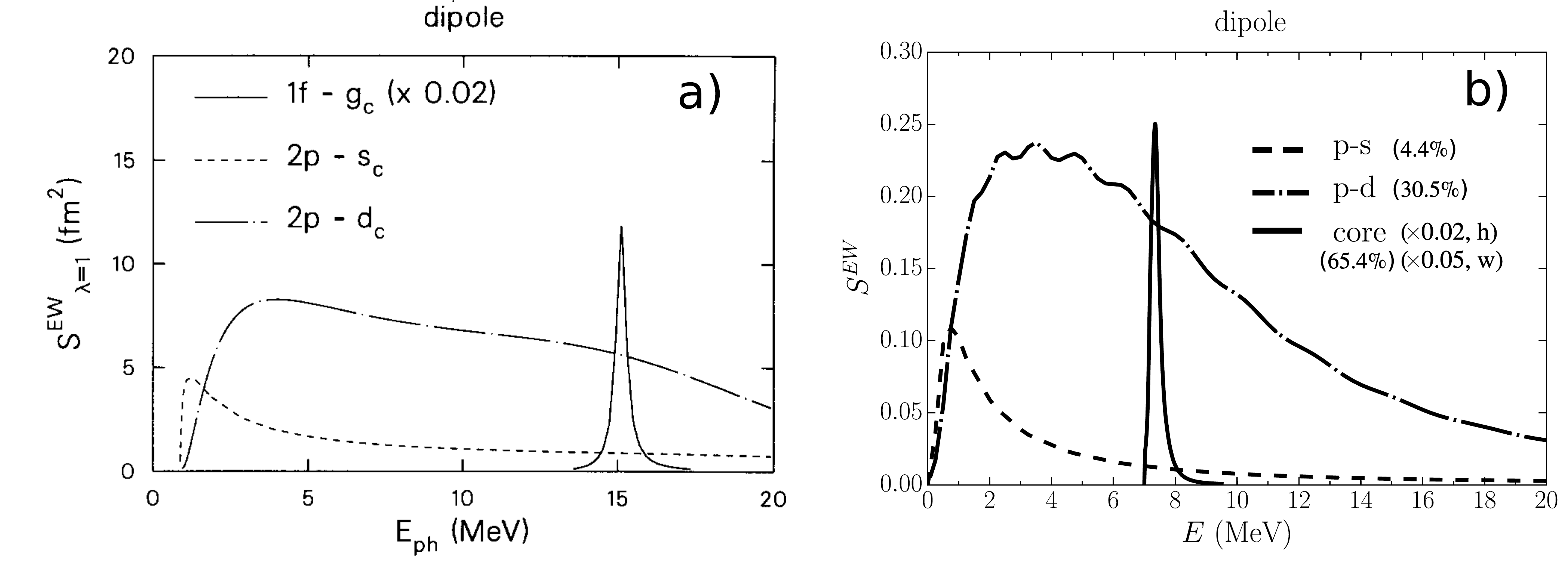}
	\end{center}
	\caption{Unperturbed contributions to the dipole EWSR for a single type of nucleons (neutrons), from selected transitions into the continuum as indicated by the labels; {\bf a)} after Fig. 8 of ref. \cite{Catara:96}; {\bf b)} present work (note the scaling factors in both height (h) and width (w)); the \% associated with the $p-s$, $p-d$ halo particle--hole ($ph$) contributions and with the core $ph$ ones (quantities in parenthesis), sum to 100\% as expected (see Eq. (\ref{eq:S}))}     
\label{fig:4}
\end{figure}

In Fig. \ref{fig:4} a) we reproduce for convenience (the upper part of) Fig. 8 of \cite{Catara:96}, which displays contributions to the dipole EWSR $S^{EW}$ (see Eq. (\ref{eq:S}) below), in comparison with our results for the neutron degrees of freedom. (Fig. \ref{fig:4} b) ), in keeping with the fact that in \cite{Catara:96} only one type of nucleons are considered, and that no explicit mention of  a Coulomb potential is made.
Aside from the fact that the high lying peak is, in our case, broader than in that of the example of  \cite{Catara:96}, in keeping with the important role played by Landau damping in the small system $^{11}$Li, the similitude of both results is apparent: low--lying concentration of strength in the continuum. In keeping with the fact already mentioned in Sect. \ref{subsec:mean}, that single--particle and collective motion emerge from the same features of the nuclear forces \cite{Mottelson:62}, this phenomenon is at the basis of the mechanism to generate unperturbed ($ph$) neutron skin (halo) transition densities behaving differently (out of phase) than unperturbed ($ph$) core ones, as discussed in Sect. \ref{subsec:mechanism}. And because of the small overlap between 
the core single-particle states and those at threshold the associated, new, uncorrelated dipole strength will mainly remain at low energy even after the residual 
dipole interaction has been switched on (QRPA).

Within this context let us now turn to the dipole ($\lambda = 1$) energy weighted sum rule for only neutrons (N), namely
\begin{equation}
 S^{EW} = \frac{9}{4\pi} \frac{\hbar^2}{2m} N.
 \label{eq:S}
\end{equation}
Consequently, whatever strength is found at low--energy cannot be but at the expenses of the high--lying  peak, as is shown by the percentages of $S^{EW}$ we display in Fig. \ref{fig:4} b).

\begin{figure}[ht!]
	\begin{center}
		\includegraphics[width=0.75\textwidth]{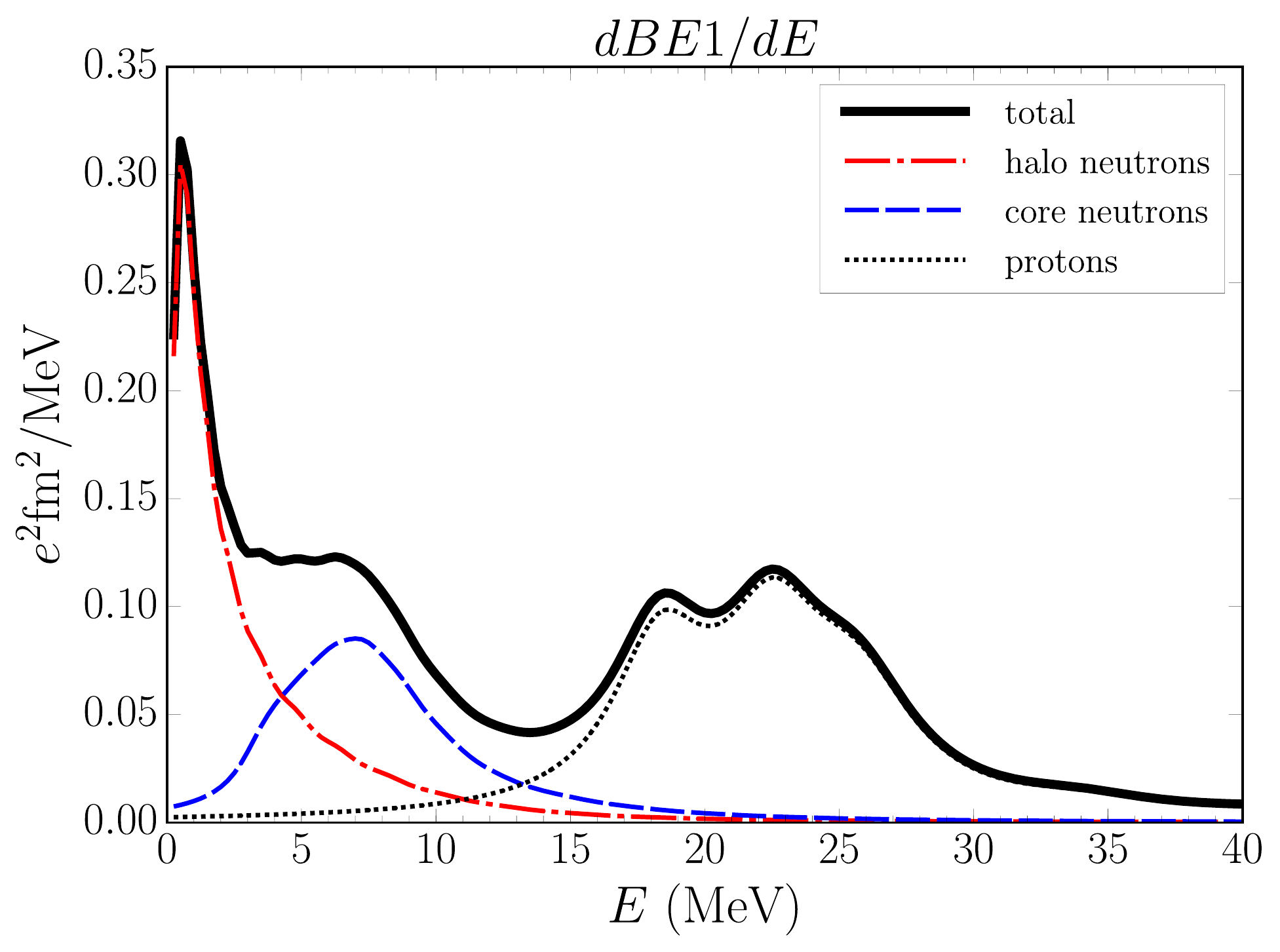}
	\end{center}
	\caption{Total (protons plus neutrons) unperturbed contributions to the dipole strength function. Integrated over energy $E$, it accounts for 100\% of the EWSR:
	$S(E1) = \frac{9}{4\pi}\frac{\hbar^2 e^2}{2m}\frac{NZ}{A}$ \cite{Bohr:75}}     
\label{fig:5}
\end{figure}

If one now considers also the protons, the full unperturbed dipole function of $^{11}$Li acquires an accumulation of strength around $20-25$ MeV, precursor of the GDR which again emphasizes the (single--particle)--collective motion connection (Fig. \ref{fig:5}). And this concludes the discussion of the unperturbed ($ph$) dipole response function.

We then diagonalize in the corresponding basis of two quasiparticles states the dipole interaction in the QRPA as explained in \cite{Barranco:01}. Adjusting the value of the strength around the self--consistent estimate \cite{Bohr:75} so as to have the lowest root at zero energy (spurious state, see below), one obtains the energies and wavefunctions of the QRPA normal modes, that is, the correlated linear dipole function. Examples of these quantities are given in Tables I and II.

Making use of these wavefunctions, the associated transition densities $\delta \rho^i$, providing a compact visualization of the spatial structure of the normal modes $i$, in particular their proton and neutron relation and thus isospin character, can be calculated. Before proceeding with the discussion of the results, let us dwell on a technical point.

\begin{figure}[ht!]
	\begin{center}
		\includegraphics[width=0.75\textwidth]{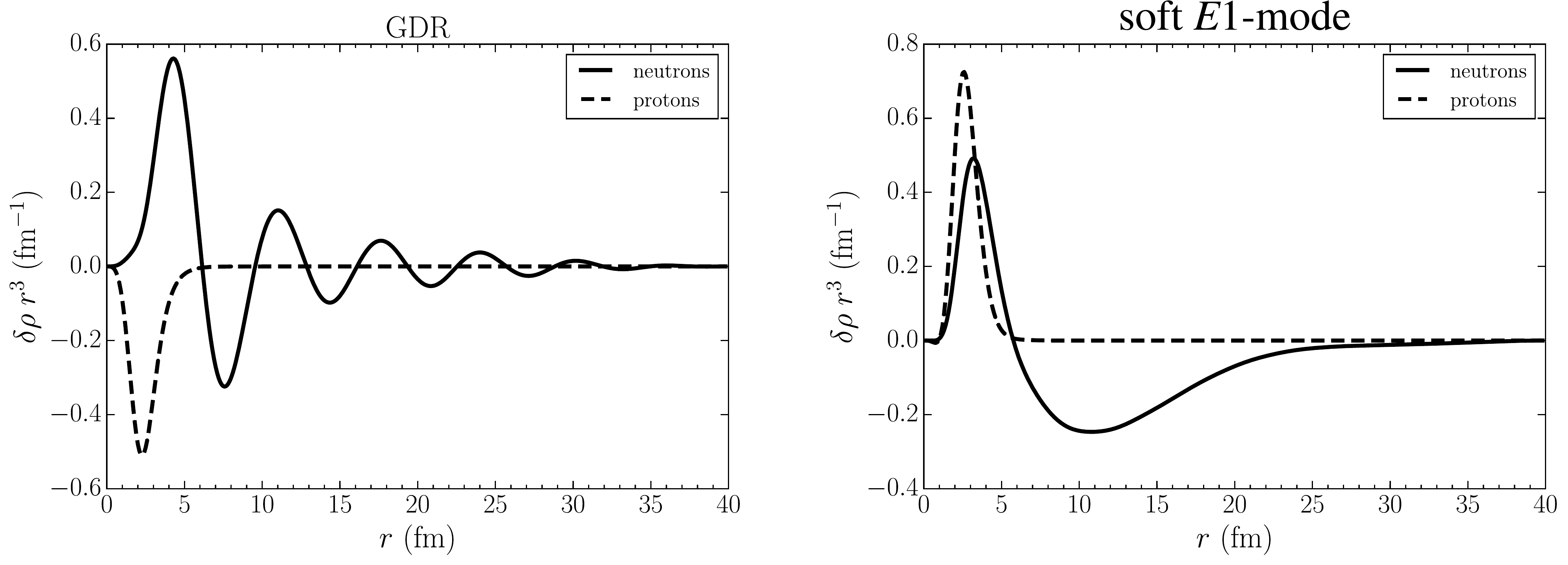}
	\end{center}
	\caption{Transition densities multiplied by $r^3$ associated with three states representative of the soft $E1$-mode and with the GDR of $^{11}$Li, calculated making use of the wavefunctions displayed in Tables I and II. }     
\label{fig:6}
\end{figure}

Limitations in numerical accuracy results in the fact that the lowest root of the QRPA solutions, labeled in what follows {\it closest to zero}  (ctz), is at a very small but finite energy. A limitation which results in the presence of some (small) amount of spuriosity in the RPA solutions ($\widetilde X$, $\widetilde Y$) and transition densities. In order to eliminate them, we have followed a method similar to that of ref. \cite{Hamamoto:98}. Namely, that of subtracting an amount $a^i$ of the transition density associated with that of the (ctz) mode, from all the QRPA roots $i \neq $(ctz), that is
\begin{equation}
 \delta \rho^i_p = \delta \widetilde \rho^i_p - a^i_p \delta \widetilde \rho^{\textrm{ctz}}_p
\end{equation}
and
\begin{equation}
 \delta \rho^i_n = \delta \widetilde \rho^i_n - a^i_n \delta \widetilde \rho^{\textrm{ctz}}_n.
\end{equation}
The coefficients $a^i$ are determined through the relations (see Fig. \ref{fig:6})
\begin{equation}
 \int \delta \rho^i_p r^3 \textrm{d}r + \int \delta \rho^i_n r^3 \textrm{d}r = 0,
\end{equation}
and
\begin{equation}
 \int \delta \rho^i_p r^3 \textrm{d}r = N/A \int \delta \rho^i_p r^3 \textrm{d}r - Z/A \int \delta \rho^i_n r^3 \textrm{d}r,
\end{equation}
so as to ensure center of mass rest and isovector strength conditions fulfilled.

% Making use of these wavefunctions, one has calculated the proton--neutron transition densities (see e.g. Fig. \ref{fig:6}) fulfilling
% \begin{equation}
% \int \delta \rho_n r^3 \textrm{d}r + \int \delta \rho_p r^3 \textrm{d}r = 0,
% \label{eq:cm}
% \end{equation}
% for each individual state (center of mass at rest; concerning the complete elimination of the Goldstone mode from the wavefunction see App. B). 

\begin{figure}[ht!]
	\begin{center}
		\includegraphics[width=0.75\textwidth]{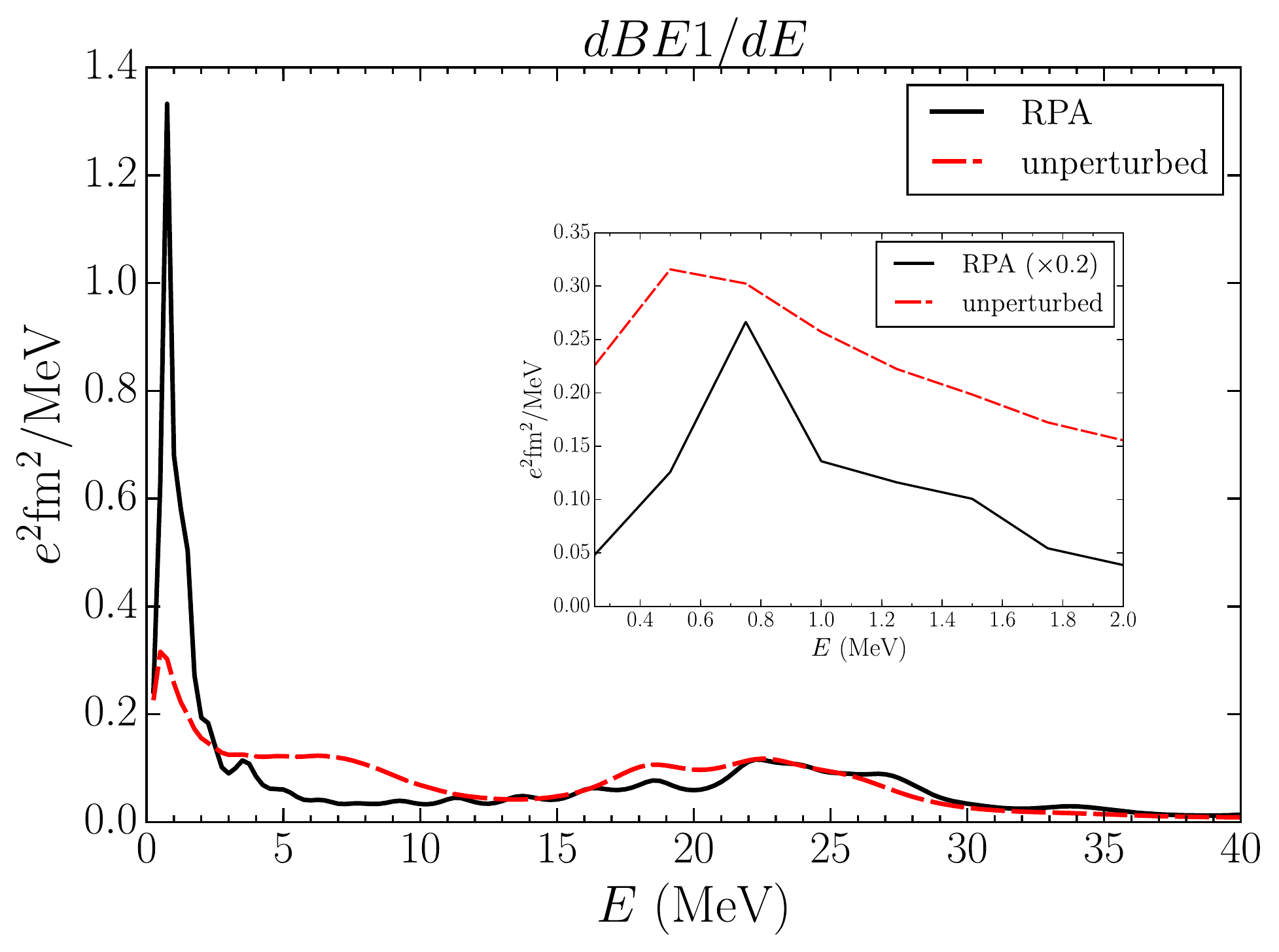}
	\end{center}
	\caption{Full response function $\textrm{d}B(E1,E)/\textrm{d} E$ (e$^2$fm$^2$/MeV)
	associated with $^{11}$Li. In the inset the $\textrm{d}B(E1,E = 0-2$ MeV)$/\textrm{d}E$ (e$^2$fm$^2$/MeV) is again displayed. }  
\label{fig:7}
\end{figure}

Making use of these quantities, the associated $B(E1)$--values were worked out. Broadening each individual state with a Lorentzian function of width (FWHM) $\Gamma$ which changes slowly with energy (square root behaviour and equal to $0.25$ MeV ($E = 1$ MeV) and $1.10$ MeV ($E = 20$ MeV)), the corresponding $dB(E1,\omega)/d\omega$ (e$^2$fm$^2$/MeV) strength function (Fig. \ref{fig:7}) and associated EWSR as a function of energy (Fig. 8) were worked out. One can observe two concentrations of strength. The lowest one (soft dipole mode) built up of approximately $1.5 \times 10$ states, the highest (GDR) containing about $1.5 \times 10^2$ states. The soft $E1$--mode is characterized by a centroid $E_{soft} \approx 0.75$ MeV, a width (FWHM) $\Gamma_{soft} \approx 0.5$ MeV and EWSR of 6 \%. 
The fact that the wavefunctions of this mode have about twenty components with $|X|>0.1$ and that $\sum |Y|^2 \approx 0.4$ (ground state correlations), testifies to the robust collectivity of the soft $E1$--mode.
The second peak displays: $E_{GDR} \approx 24$ MeV, $\Gamma_{GDR} \approx 11$ MeV, and EWSR $\approx 90\%$. It is of notice that, although in the inset to Fig. \ref{fig:7} we display, for the sake of clarity, the soft $E1$--mode in the energy range $0-2$ MeV, this state extends up to $6-7$ MeV (see Fig. 8a)). 

Aside from the difference in energies and widths, the two collective modes also display, from a structural point of view, rather different dynamical properties. In the case of the GDR (Fig. 9 (b)) the neutron and the proton transition densities show (disregarding local oscillations) rather similar shapes which are in anti--phase (protons vibrating against the neutrons). 
%The minimum of the proton transition density essentially coincides with the radius of the bare single--particle potential ($R_0 = 2.8$ fm), while the maximum of the neutron transition--density is found at $\approx 4.8$ fm, and the first sign change of $\delta \rho_n$ is observed at $\approx 6.8$ fm. One can then estimate the neutron skin to be $\approx 2$ fm (4 fm).
Concerning the soft dipole mode (Fig. 9(a)), the transition density shows an in--phase motion of the neutrons and protons belonging to the inner part of the nucleus (the core region)
which is  in  anti--phase with the neutron (halo) skin. 
Within this context, it is of notice the similitude of the present results with those associated with the PDR of $^{208}$Pb and of $^{122}$Sn and, reported in 
Fig. 3 of ref. \cite{Ryezayeva:02} ($^{208}$Pb) and in Fig. \ref{fig:5} of ref. \cite{Tsoneva:08} ($^{122}$Sn), as can be seen also from Fig. 9 (right and left panels labeled $^{208}$Pb a $^{122}$Sn respectively). A similitude which extends, as expected, also to the transition densities associated with the GDR and displayed in the same figure.

Further support for the validity of the mechanism unifying the physics at the basis of PDR and soft $E1$--modes is provided by studies of the nuclear dipole response at finite temperature. The phenomenon of thermally unblocking particle--hole pairs ($\widetilde{ph}$) with both single--nucleon states above the Fermi energy allows for the appearance of low--energy dipole states extending outside the core nucleons and leading to a neutron skin which oscillates out of phase with respect to a core in which protons and neutron 
vibrate  in phase.
Such a behaviour, typical of a PDR, is observed e.g. in connection with the $E_X = 3.49$ MeV dipole state of the neutron rich nucleus ${}^{68}_{28}$Ni$_{40}$ excited at a temperature of $T=3$ MeV. as testified by the transition density displayed in Fig. 10, and of the corresponding wavefunction made out mainly of $\widetilde{ph}$ components with $\varepsilon_{\widetilde{ph}} > \varepsilon_F$, namely $(3s_{1/2} \rightarrow 3p_{3/2})n$, $(2d_{5/2} \rightarrow 3p_{3/2})n$ and $(3s_{1/2} \rightarrow 3p_{1/2})n$ (see Table I of \cite{Wibowo:18}). Increasing the temperature to $T=4$ MeV, the strongest dipole state lying below 10 MeV, and carrying 72.4\% of the 0--10 MeV strength, moves down in energy to $E_x = 2.55$ MeV.
The robustness of the above parlance is further confirmed by results found in the case of the neutron deficient nucleus ${}^{100}_{50}$Sn at $T=3$ MeV, 
displaying in this case  a proton skin and, fittingly, a soft $E1$--mode (e.g. the state at 4.13 MeV), where the proton skin oscillates against an isospin saturated core \cite{Litvinova:18}.

In an attempt at shedding light on the role quantal fluctuations play in the E1-soft mode of $^{11}$Li under discussion, we have recalculated the transition densities, but this time without including ground state correlations (GSC). That is, setting the $Y$-- component of the QRPA wavefunctions to 0, and normalizing to 1 the sum squared of $X$ components. The results are displayed in Fig. 11. In the case of the soft dipole mode the effect of GSC is quite important and constructively coherent throughout for both protons and neutrons (Fig. 11(a)). In the case of the GDR, ground state correlations, which essentially only act on the core neutrons, lead to destructive interference (Fig. 11(b)).  Eliminating their effect results in the elimination 
of oscillations in $\delta \rho_n$ ($r \lesssim R_0$), and lead to a transition density which better resembles the paradigm of an isovector mode. A result consistent with the fact that the $Y$--components foster, in the present case ($^{11}$Li),  the ($pp$) component of two--quasiparticle collective modes, and that the GDR is essentially a correlated ($ph$)--excitation. In other words, GSC oppose the collectivity of the GDR. By the same token one can posit that the two--quasiparticle wavefunction of the soft $E1$--mode is dominated by its ($pp$)--component, consistent with a vortex--like mode. Expressing it differently, a dipole Cooper pair (see e.g. \cite{Ryezayeva:02, Bertsch:94} and refs. therein).

Summing up: a) the concentration of 6\% of the dipole EWSR in a peak of centroid {$E_X \lesssim 1$ MeV} and width 
$\Gamma \approx 0.5$ MeV; b) the wavefunctions of the states forming it having  about 15 phase correlated components, displaying about 30\% of ground state correlations, and associated transition densities consistent with a well developed neutron skin ($\Delta r_{np} \approx 1.71$ fm) which oscillates out of phase with 
respect to an isospin saturated core, testifies to the fact that the soft $E1$--mode of $^{11}$Li, 
%but also similar excitations observed in other exotic nuclei close to the neutron dripline, 
can be viewed as an elementary mode of excitation.
% well deserving a name: namely that of Pygmy Dipole Resonances (PDR).

There is however, an essential difference between the otherwise physically similar modes. In e.g. $^{208}$Pb and $^{122}$Sn although the vibrational amplitude of the PDR and the role of pairing are larger than in the case of the GDR, the PDR is still a small amplitude mode. On the other hand pairing, at the level of single Cooper pair, plays an important role in the case of the low--energy dipole resonance of $^{11}$Li, being this mode a large amplitude vibration.

Within this context, the analysis of the flow pattern associated with low-energy E1-modes (see e.g. \cite{Repko:13} as well as \cite{Ryezayeva:02}) may provide 
new possibilities to better characterize the broad species low-energy E1-modes, eventually distinguishing between e.g. the low- and high-energy 
bin (6.0-8.8 MeV and 8.8 MeV-10.5 MeV in $^{208}$Pb) and these, from the soft-E1 mode of $^{11}$Li. A possibility to help at making  the conclusions
of \cite{Repko:13} non only a theoretical classification, is that of of probing these modes through both anelastic processes (e.g. $\gamma,\gamma'$,
inelastic scattering, Coulomb excitation, etc) and two-particle  transfer reactions (e.g. (p,t), (t,p) etc), and to study the role ground state correlations
play in reproducing (predicting) the associated absolute differential cross sections. A subject touched upon in connection with Fig. 11.

\section{Hindsight} \label{sec:hindsight}

The origin of the pigmy resonance in light halo nuclei, namely, parity inversion, is a many-body effect going beyond mean field as testified by the large 
particle--vibration coupling vertices between the quadrupole vibration of the core and the halo neutrons $s_{1/2}$ ($\approx -3 $ MeV) and $p_{1/2}$ ($\approx$ -3.9 MeV) and by the population of the lowest $1/2^-$ member of the multiplet $(2^+ \otimes p_{3/2}$) of $^9$Li in the reaction $^{11}$Li(p,t)$^9$Li ($1/2^-$; 2.69 MeV) \cite{Tanihata:08,Potel:10} and of the 
quadrupole vibration in $^{10}$Be in the reaction $^{11}$Be(p,d)$^{10}$Be(2$^+$; 3.33 MeV) \cite{Winfield:01,Barranco:17}. Let alone by the results 
of the QRPA calculation presented here, namely of the GDR, 
of the soft $E1$--mode  and of a dipole state at zero energy, comprehensively exhausting the TRK sum rule, these last two states 
carrying 6\% ($\approx 1 B_W (E1))$ and 0\% of it respectively)  (within this context, see \cite{Hamamoto:98,Sagawa:01}). 

Origin also documented by the experimental values of the inelastic $^{11}$Li(d,d')$^{11}$Li($1^-$; 1 MeV) \cite{Kanungo:15} and $^{11}$Li(p,p')$^{11}$Li($1^-$; 0.80 MeV) \cite{Tanaka:17} cross sections, 
as well as of the decay strength $B(E1; 1/2^- \to 1/2^+)$
between the parity inverted first excited $|\widetilde{1/2}^- \rangle$ (0.18 MeV) and the ground state $|\widetilde{1/2}^+ \rangle$ states of $^{11}$Be \cite{Kwan:14}, both quantities corresponding to
$\approx $ 1 $B_W(E1)$ ($\approx 0.1$ e$^2$ fm$^2$). Within this context it is important to remember the subtle relation existing between collective and single-particle 
motion. Assuming the full TRK sum rule $S(E1)= 14.8 \frac{NZ}{A}$ e$^2$ fm$^2$ to be concentrated in the GDR ($\hbar \omega_{GDR} \approx $ 80 MeV A$^{-1/3}$), one
expects the associated E1-strength to be  
\footnote{In this estimate we have used the approximate relation valid for nuclei with large neutron excess like $^{208}$Pb and $^{11}$Li, $N/A \approx 0.67$ and 
$Z/A \approx 0.33$, leading to $NZ/A\approx 0.2 $A.}
$\approx 3.75 \times 10^{-2} A^{4/3}$ e$^2$ fm$^2$, which for A=11 amounts to  0.92 e$^2$ fm$^2$ and thus to $\approx 9 B_W(E1)$. Let us now calculate 
the $B(E1)$ between the dressed  single-particle states of $^{11}$Be 
 $| \widetilde{p_{1/2}}>$ and $|\widetilde{s_{1/2}}>$. Taking into account contributions arising from many-body processes associated quadrupole, octupole
 and pairing  vibrational modes  as explained in \cite{Barranco:17} (see supplemental material) 
 one obtains $\approx 0.12$ e$^2$ fm$^2$, i.e. almost 13\% of the 
E1-strength associated with a GDR assumed to contain 100\% of the EWSR. Whether one calls this result a collective or a pure single-particle transition is arguably just semantics. 
%The dressing of the $s_{1/2}$ and $p_{1/2}$ states by the quadrupole mode leading to $|\widetilde{1/2}^+> = \sqrt{0.80} |s_{1/2}> + \sqrt{0.2} |(d_{5/2} \otimes 2^+)_{1/2^+}$ and
%$\widetilde{1/2}^-= \sqrt{0.84} |p_{1/2}> + \sqrt{0.16} |(p_{1/2} \otimes p_{3/2})_{2^+} \otimes 2^+)_{0^+}, p_{1/2}>$ leads to a value $B(E1) \approx 0.8 \times 0.84 \times 0.17$ e$^2$ fm$^2$ =
%0.11 e$^2$ fm$^2$ consistent with the reduction of single-particle  content in the initial and final states. Including the coupling to the octupole and pairing vibration  results in 
%$B(E1)  \approx$   0.12 e$^2$ fm$^2$ \cite{Barranco:17}. 

The usual depletion of the  low-energy E1 strength measured by the dipole effective charge squared $(e(E1)_{eff})^2 \approx (-0.5 \tau_z (1+\chi))^2
\approx 10^{-2} $ e$^2$ ($\chi = -0.7, \tau_z = \pm 1 $ (n,p)) is here essentially screened out by the poor overlap between core and halo nucleons. 
In other words. one 
finds in the case of $^{11}$Li a negligible depletion instead of the conspicuous  one 
% leading to a final value 
%$B(E1; \tilde{1/2^+} \to \tilde{1/2^-}) = 0.11 $ e$^2$ fm$^2$, i.e. to only a 8\% depletion {\color{green} rispetto a cosa? Noi non teniamo conto dell'effective charge nel calcolo}.  This, instead of the conspicuous depletion 
observed in  stable nuclei lying 
along the stability valley and leading to typical E1-single particle strength $\leq 10^{-2}$ $B_W(E1)$ (A $\geq 45$) (see \cite{Endt:79,Endt:81,Endt:93,Martin:07}).

A second issue concerning the dipole response in nuclei with neutron excess is related to the isovector-isoscalar character of the collective mode. In a similar way in which the states of a nucleus 
in a very strong external magnetic field pointing along the $z-$axis cannot be characterised by a definite angular momentum, the isoscalar  and isovector states get strongly  
mixed in nuclei like $_3^{11}$Li$_8$ displaying very large neutron excess ((N-Z)/A $\approx$ 0.5).

\section{Conclusions}\label{sec:conclusions}
Weakly bound two quasiparticle ($ph$) states with low centrifugal barriers, e.g. ($sp^{-1}$) configurations at threshold are, in neutron rich nuclei, at the basis of the presence of a neutron skin. Thus, of collective modes in which the neutron skin oscillates against the nucleons of the core. Both in nuclei lying along the stability valley 
as well as  in exotic halo systems lying along the neutron dripline, these vibrations display common, universal features and rightly deserve a common name. 
Likely that of Pygmy Dipole Resonances (PDR).

Halo Cooper pairs or better halo pair addition vibrations and pygmy dipole resonances are two novel plastic modes of nuclear excitation. Experimental studies of these excitations, in particular of pygmy resonance based on excited neutron halo states are within reach of experimental ingenuity and techniques. They are expected to shed light on a basic issue which has been with us since the formulation of BCS theory of pairing: the microscopic mechanism to break gauge invariance and thus of the variety of origins of nuclear pairing, beyond the bare $NN$-pairing force.

Furthermore, they are likely to open a new chapter in the probing of the Axel-Brink hypothesis: its state dependence. This phenomenon can be instrumental in modulating the transition between cold and warm (equilibrated) excited nuclei, let alone provide a microscopic way to study a new form of inhomogeneous damping. \textit{Namely radial isotropic deformation. The importance of this mechanism, which has partially entered the literature under the name of neutron-skin}, is underscored by the fact that, in $^{11}$Li
 it is able to bring down from the GDR (or equivalently, impede the GDR to bring up continuum-like strength) by tens of MeV, a consistent fraction of the TRK sum rule associated with the GDR. 

But even more important, because it provides a laboratory to study  the fashion in which the above mentioned ``deformation'' affects nuclear matter. Matter which is little compliant to undergo either compressions (saturation) or ``depressions''. In the first case by reacting through a mini supernova. In the second, by obliterating the effects of the short range strong force acting in the $^1S_0$ channel and of the symmetry potential. 

\section{Acknowledgements}\label{sec:acknowledgements}
Discussions with H. Lenske are gratefully acknowledged.
F. B. and E. V. acknowledge funding from the European Union Horizon 2020 research and innovation program under Grant Agreement No. 654002. 
F.B. thanks  the Spanish Ministerio de  Econom\'\i a y Competitividad and FEDER funds under project FIS2017-88410-P.
During the inception of the present work A.I. was supported by  the  Helmholtz  Association  through  the  Nuclear  Astrophysics  Virtual  Institute
(VH-VI-417) and the Helmholtz International Center for FAIR within the framework of the LOEWE program launched by the state of Hesse.
We gratefully acknowledge technical help from Davide S. R. Azzini. \clearpage

\appendix

\section{Inhomogeneus damping of PDR in anomalously radially extended exotic halo nuclei}\label{app:inhomogeneus}
The correlation length of the Cooper pair associated with the two halo neutrons of $^{11}$Li is (\cite{Schrieffer:64}, p. 18),
\begin{equation}
 \xi_0 = \frac{\hbar v_F}{\pi |E_{corr}|} \approx 20 \textrm{ fm},
\end{equation}
in keeping with the fact that\footnote{One can write $(v_F/c)=(k_F)_{fm^{-1}}/5$. For nuclei lying along the stability valley, $k_F\approx1.36\textrm{fm}^{-1}$ (\cite{Bohr:69}, p. 140). An estimate for $^{11}$Li is provided by $k_F(^{11}\textrm{Li})\approx1.36\textrm{fm}^{-1}(R_0(^{11}\textrm{Li})/R(^{11}\textrm{Li}))\approx0.8\textrm{fm}^{-1}$. Thus $(v_F/c)\approx0.16$ and $\epsilon_F\approx13$ MeV.} in $^{11}$Li $v_F/c \approx 0.16$ and $E_{corr} \approx - 0.5$ MeV.

Making use of the radius of $^9$Li obtained from systematics, $R_0 = 1.2 \times 9^{\frac{1}{3}} \textrm{ fm} = 2.5 \textrm{ fm}$, one can calculate the effective radius of $^{11}$Li using the correlation length of a Cooper pair,
\begin{equation}
 (R^2_{eff} (^{11}\textrm{Li}))^{1/2} = \left( \frac{9}{11} R^2_0 (^{9}\textrm{Li}) + \frac{2}{11} \left( \frac{\xi_0}{2} \right)^2 \right)^{1/2} = 4.8 \textrm{ fm}; \quad (\langle r^2 \rangle^{1/2} = \sqrt{\frac{3}{5}} R_{eff} \approx 3.7 \textrm{ fm}),
 \label{eq:Reff}
\end{equation}
not inconsistent with the experimental findings $ \langle r^2 \rangle^{1/2} = 3.55 \pm 0.1 \textrm{ fm}$ \cite{Kobayashi:89}.

One can then estimate the single-particle overlap probability associated with $\varphi(^{11}$Li$)_{halo}$ and with $\varphi(^{11}$Li$)_{0}$, making use of Eq. (\ref{eq:Reff}) and of the value of the radius (obtained from systematics) of an hypothetical normal nucleus of mass $A=11$, namely $R_{0} (^{11}\textrm{Li}) = 1.2 \times (11)^{\frac{1}{3}} \textrm{ fm} \approx 2.67 \textrm{ fm}$, i.e.
\begin{equation}
 \mathcal{O} = |\langle \varphi(^{11}\textrm{Li})_{halo} | \varphi (^{11}\textrm{Li})_0 \rangle|^2 =
          \left( \frac{R_{0}}{R_{eff}}\right)^3 =
          \left( \frac{2.67}{4.8} \right)^3 \approx 0.17.
\end{equation}

Thus, the sloshing back and forth of the neutrons against the protons associated with the core leads to a GDR (consistent with the {$R_0 = 1.2 A^{1/3}$ fm} systematics), while the transition densities of the PDR are  dominated by the neutron halo contributions with respect to an isospin saturated core.
% against the protons gives rise to the pygmy resonance, a plastic, large amplitude dipole collective mode based on an exotic, highly extended, halo ground state $|gs(^{11}\textrm{Li})\rangle$. 
% Therefore, one can posit that halo states based on $s$-- and $p$-- orbitals at threshold, (halo anti-pairing effect, cf. \cite{Bennaceur:00,Hamamoto:03,Hamamoto:04}) whether ground or excited states, will carry on top of it a pygmy resonance.

We are likely confronted with both a new mechanism to break gauge invariance \cite{Cooper:11a}, as well as a novel way to study the evolution of giant resonances based on excited states \textit{and thus learn about homogeneous radial deformation of both ground and excited states}.
 
Within this context one can mention the extremely different role GQR and low-lying large amplitude quadrupole vibrations play in the fission and exotic decay of nuclei. The low-lying vibrations are essential in bringing the system into the necking situation (cf. \cite{Brink:05} Ch. 7, and refs. therein), giant quadrupole resonances hardly being able to contribute other than with polarization effects.

\section{Continuum discretization}

%Concerning the so-called continuum effect introduced in ref. \cite{Catara:96} (see also \cite{Nagarajan:05}),
%we note that in the case of the dipole mode, the associated TRK sum rule depends only on the number
%of nucleons and not on the radius. Consequently, the fact that the PDR of $^{11}$Li carries 7\% of the EWSR, is a consequence
%of the redistribution of the high-lying dipole response (GDR) expected in the case of nuclei lying along the 
%stability valley, as resulting from the calculation of the full dipole response function in QRPA. 

There  are different  possibilities to simultaneously treat bound and continuum
states. Among others, to use a harmonic oscillator basis properly adjusting the restoring force constant  \cite{Calci:16}, to discretize the continuum  by enclosing the 
mean field potential in a spherical box  of appropriate  radius to guarantee convergence  \cite{Bertsch:91} (see also \cite{Mazur:17}), let alone a detailed treatment of it in terms 
of running waves, as is done in describing reaction processes (se e.g. Sect .IV.1), or using Lorenz integral techniques \cite{Bacca:14}. 
%In any case, there is one, a priori, condition any of the corresponding methods 
%must fulfil. The integral over the whole space (volume) of the sum modulus squared of the single-particle wavefunction must be equal to the number of particles.
%Said it differently, the maximum amount of energy a system can absorb from a photon has a well defined value (see e.g. \cite{Broglia:05c} and refs. therein). Value controlled in turn
%by the simple fact that the double commutator of the dipole operator and the kinetic energy is independent of the correlations of the many-body system under study, 
%and only depends in the mass of the particles forming it. 

Different methods to simultaneously treat bound and continuum states have been used  in the calculations of  the low-lying  dipole response.
We have  discretized the continuum  by enclosing a Saxon-Woods potential in a spherical box  of appropriate  radius to guarantee convergence.
Other calculations  have adopted an exact treatment of the continuum \cite{Nakatsukasa:05,Mizuyama:09}, have used a harmonic oscillator
basis to perform shell model calculations  \cite{Sagawa:99},  
or have adopted  the complex scaling method  \cite{Kikuchi:13}.

\clearpage

\begin{figure}[ht!]
	\begin{center}
		\includegraphics[width=0.75\textwidth]{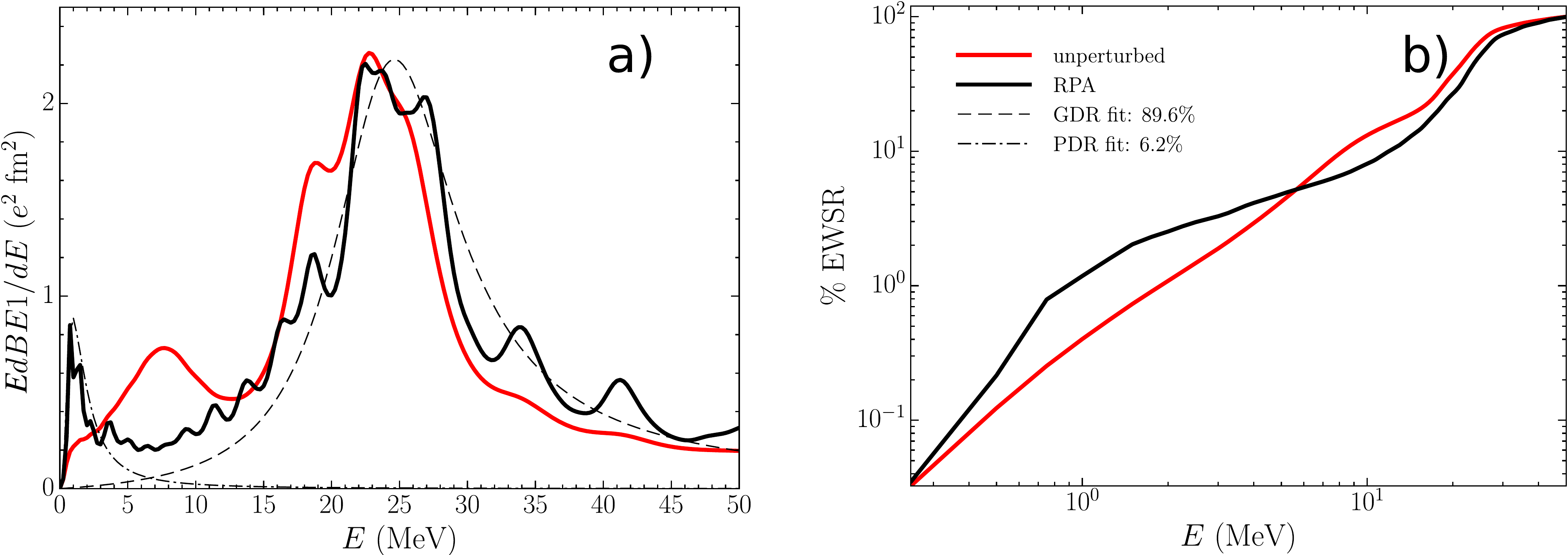}
	\end{center}
	\caption{ a) Same as Fig. 7 but multiplied by the excitation energy E.  b) Integrated EWSR as a function of the energy. }     
\label{fig:8}
\end{figure}

\begin{figure}[ht!]
	\begin{center}
		\includegraphics[width=0.75\textwidth]{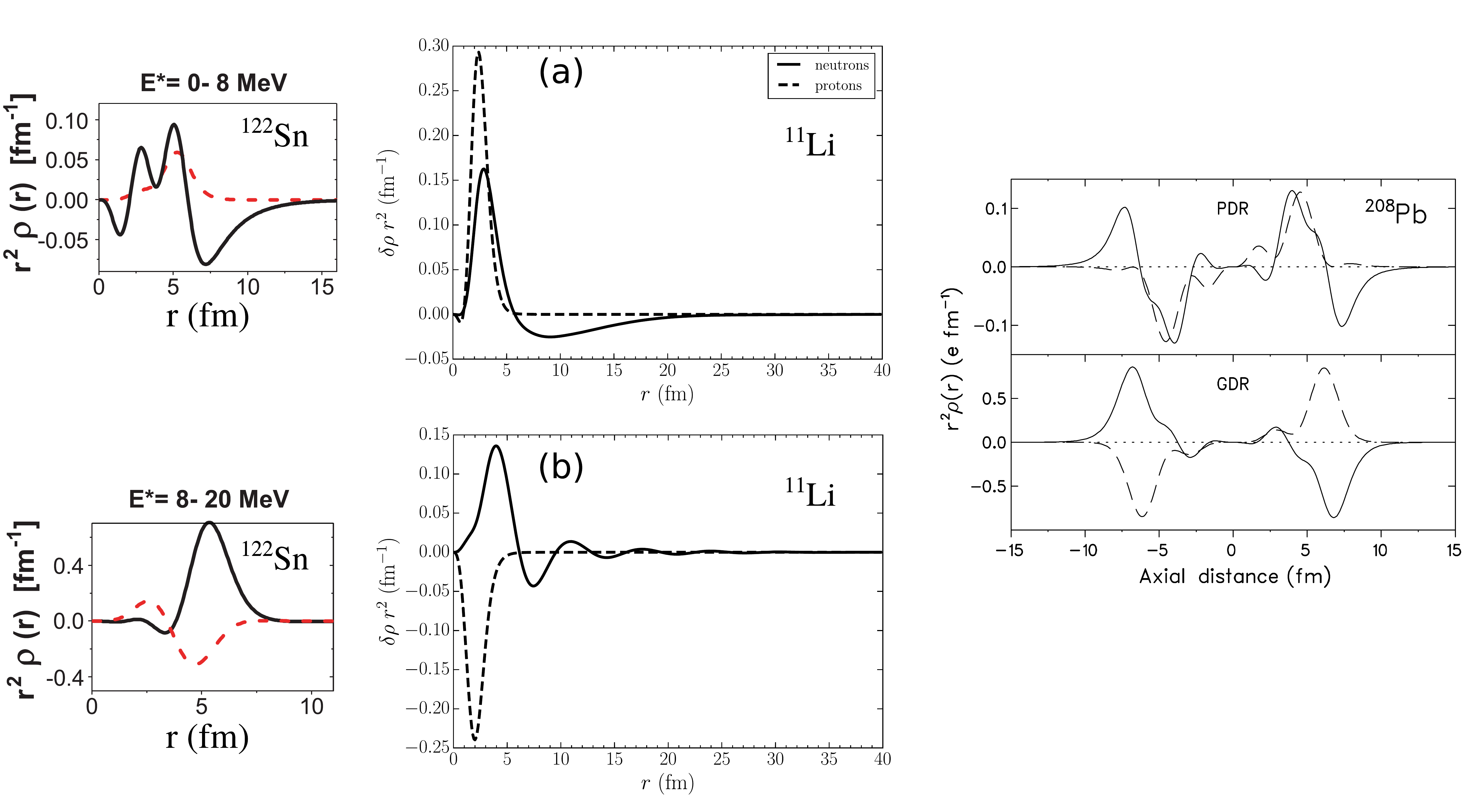}
	\end{center}
	\caption{ Transition densties multiplied by $r^2$ associated with three states representative of the soft dipole mode (a), and with the GDR (b) of $^{11}$Li (see Tables I and II). In the right and left panels the same quantities associated with the PDR and GDR of $^{208}$Pb \cite{Ryezayeva:02} and $^{122}$Sn \cite{Tsoneva:08} are displayed. }     
\label{fig:8a}
\end{figure}

\begin{figure}[ht!]
	\begin{center}
		\includegraphics[width=0.75\textwidth]{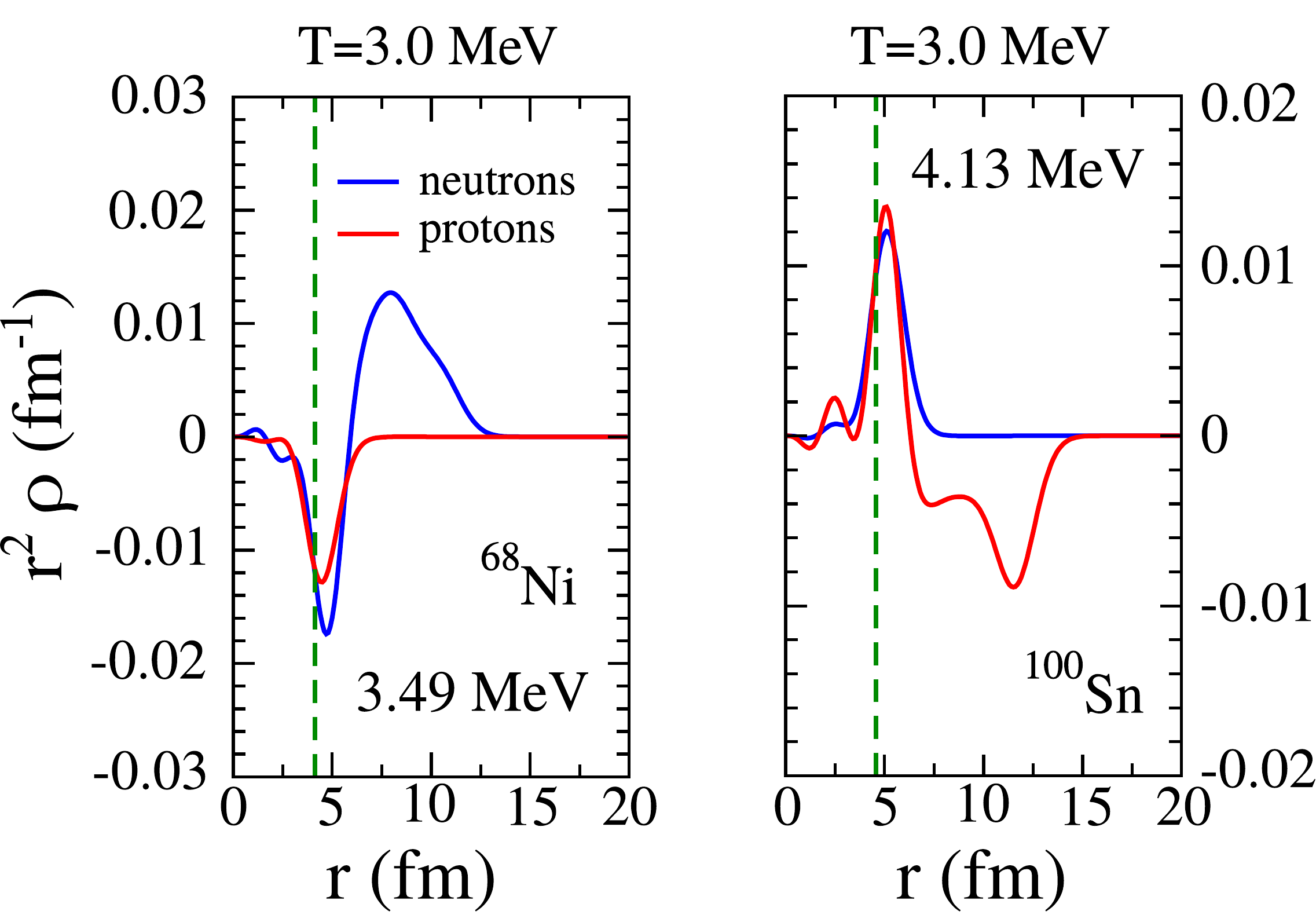}
	\end{center}
	\caption{Transition densities associated with the states at $E_X = 3.49$ MeV and 4.13 MeV of $^{68}$Ni and $^{100}$Sn respectively, at a temperature of $T=3$ MeV in both cases (after \cite{Wibowo:18} and \cite{Litvinova:18}).}     
\label{fig:9}
\end{figure} 

\begin{figure}[ht!]
	\begin{center}
		\includegraphics[width=0.75\textwidth]{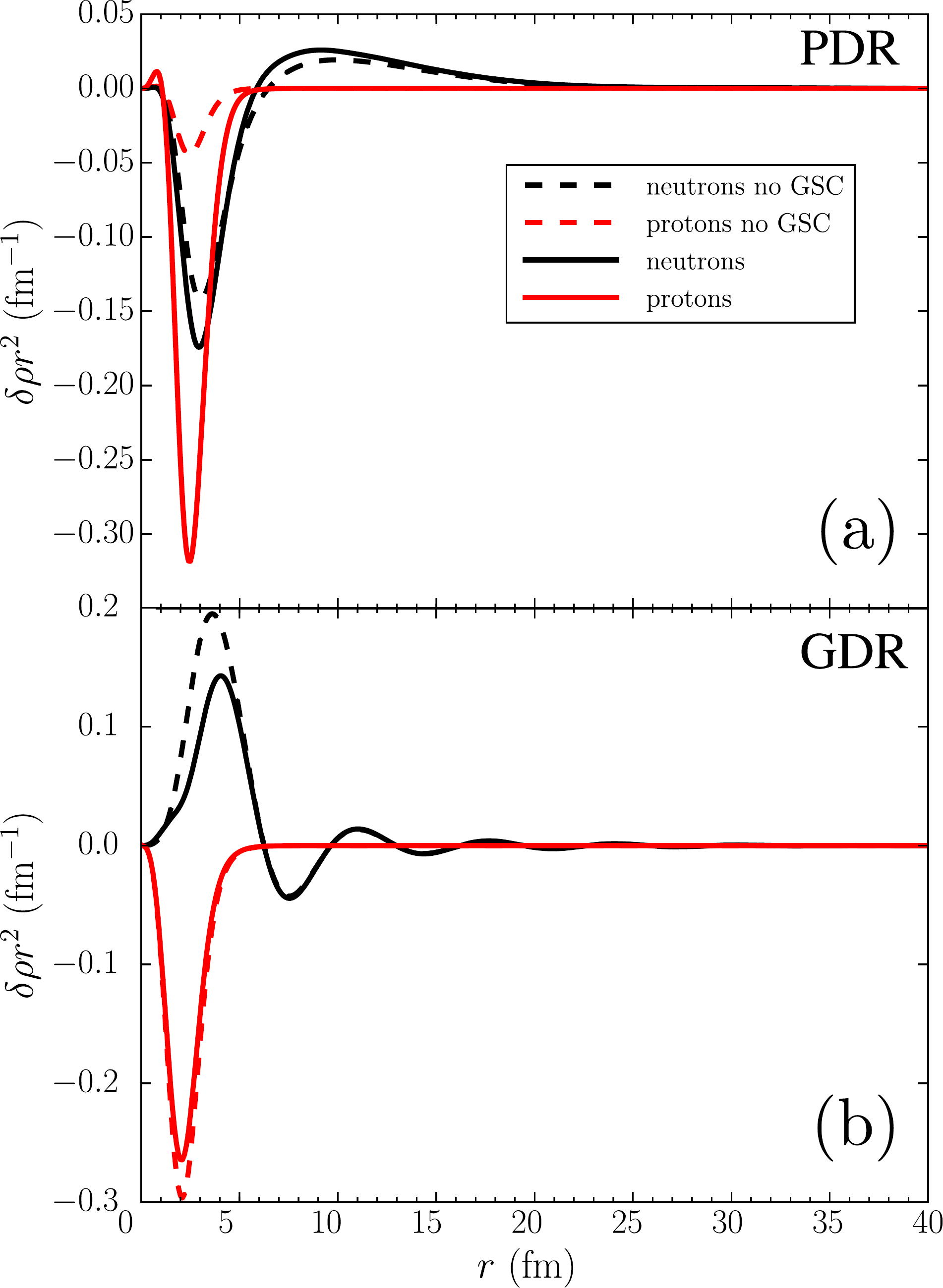}
	\end{center}
	\caption{(color online) {\bf (a)} The transition densities associated with the PDR shown in Fig. 9(a) are here displayed again in comparison with the corresponding quantities labeled (noGSC) and calculated without taking into account the Ground State Correlations (RPA continuous curve, (noGSC) dashed curve); {\bf (b)} same but for the GDR
	(Fig. 9(b)). }     
\label{fig:10}
\end{figure}

%Pygmy
\begin{table}[]
\begin{tabular}{lllll|lllll|lllll}
	% E = 300 keV                                        % E = 500 keV                                       % E = 700 keV
      & i          & j          &$X_{ij}$&$Y_{ij}$ &       & i          & j          &$X_{ij}$&$Y_{ij}$&       & i           & j         &$X_{ij}$&$Y_{ij}$ \\
         \hline
 $\nu$ & 2$s_{1/2}$ & 1$p_{1/2}$ &-0.780  & 0.078  & $\nu$ & 2$s_{1/2}$ & 1$p_{1/2}$ & -0.119 & 0.048  & $\nu$ & 3$s_{1/2}$ & 1$p_{1/2}$ & -0.118 & 0.040 \\  
 $\nu$ & 3$s_{1/2}$ & 1$p_{1/2}$ & 0.479  & 0.108  & $\nu$ & 3$s_{1/2}$ & 1$p_{1/2}$ & -0.748 & 0.074  & $\nu$ & 4$s_{1/2}$ & 1$p_{1/2}$ & -0.821 & 0.046 \\  
 $\nu$ & 4$s_{1/2}$ & 1$p_{1/2}$ & 0.220  & 0.106  & $\nu$ & 4$s_{1/2}$ & 1$p_{1/2}$ & 0.410  & 0.080  & $\nu$ & 5$s_{1/2}$ & 1$p_{1/2}$ & 0.250  & 0.046 \\  
 $\nu$ & 5$s_{1/2}$ & 1$p_{1/2}$ & 0.144  & 0.093  & $\nu$ & 5$s_{1/2}$ & 1$p_{1/2}$ & 0.181  & 0.075  & $\nu$ & 6$s_{1/2}$ & 1$p_{1/2}$ & 0.116  & 0.043 \\  
 $\nu$ & 6$s_{1/2}$ & 1$p_{1/2}$ & 0.106  & 0.080  & $\nu$ & 6$s_{1/2}$ & 1$p_{1/2}$ & 0.117  & 0.067  & $\nu$ & 1$p_{3/2}$ & 4$d_{5/2}$ & 0.144  & 0.081 \\  
 $\nu$ & 1$p_{3/2}$ & 4$d_{5/2}$ & 0.166  & 0.139  & $\nu$ & 1$p_{3/2}$ & 4$d_{5/2}$ & 0.170  & 0.121  & $\nu$ & 1$p_{3/2}$ & 5$d_{5/2}$ & 0.201  & 0.125 \\  
 $\nu$ & 1$p_{3/2}$ & 5$d_{5/2}$ & 0.241  & 0.208  & $\nu$ & 1$p_{3/2}$ & 5$d_{5/2}$ & 0.243  & 0.183  & $\nu$ & 1$p_{3/2}$ & 6$d_{5/2}$ & 0.201  & 0.135 \\  
 $\nu$ & 1$p_{3/2}$ & 6$d_{5/2}$ & 0.250  & 0.221  & $\nu$ & 1$p_{3/2}$ & 6$d_{5/2}$ & 0.249  & 0.196  & $\nu$ & 1$p_{3/2}$ & 7$d_{5/2}$ & 0.156  & 0.112 \\  
 $\nu$ & 1$p_{3/2}$ & 7$d_{5/2}$ & 0.199  & 0.180  & $\nu$ & 1$p_{3/2}$ & 7$d_{5/2}$ & 0.196  & 0.161  & $\nu$ & 1$p_{3/2}$ & 8$d_{5/2}$ & 0.113  & 0.085 \\  
 $\nu$ & 1$p_{3/2}$ & 8$d_{5/2}$ & 0.148  & 0.135  & $\nu$ & 1$p_{3/2}$ & 8$d_{5/2}$ & 0.144  & 0.122  & $\nu$ & 1$p_{1/2}$ & 9$d_{3/2}$ & -0.126 & 0.014 \\  
 $\nu$ & 1$p_{3/2}$ & 9$d_{5/2}$ & 0.110  & 0.102  & $\nu$ & 1$p_{3/2}$ & 9$d_{5/2}$ & 0.107  & 0.093  & $\nu$ & 1$p_{1/2}$ &10$d_{3/2}$ & 0.187  & 0.026 \\  
 $\nu$ & 1$p_{1/2}$ & 4$d_{3/2}$ & 0.103  & 0.075  & $\nu$ & 1$p_{1/2}$ & 2$d_{3/2}$ & 0.168  & 0.024  & $\nu$ & 1$p_{1/2}$ &11$d_{3/2}$ & 0.121  & 0.040 \\  
 $\nu$ & 1$p_{1/2}$ & 5$d_{3/2}$ & 0.119  & 0.095  & $\nu$ & 1$p_{1/2}$ & 3$d_{3/2}$ & 0.114  & 0.043  & $\nu$ & 1$p_{1/2}$ &12$d_{3/2}$ & 0.113  & 0.053 \\  
 $\nu$ & 1$p_{1/2}$ & 6$d_{3/2}$ & 0.128  & 0.108  & $\nu$ & 1$p_{1/2}$ & 4$d_{3/2}$ & 0.117  & 0.063  & $\nu$ & 1$p_{1/2}$ &13$d_{3/2}$ & 0.111  & 0.064 \\  
 $\nu$ & 1$p_{1/2}$ & 7$d_{3/2}$ & 0.128  & 0.112  & $\nu$ & 1$p_{1/2}$ & 5$d_{3/2}$ & 0.126  & 0.081  & $\nu$ & 1$p_{1/2}$ &14$d_{3/2}$ & 0.104  & 0.068 \\  
 $\nu$ & 1$p_{1/2}$ & 8$d_{3/2}$ & 0.117  & 0.106  & $\nu$ & 1$p_{1/2}$ & 6$d_{3/2}$ & 0.131  & 0.094  & $\pi$ & 1$p_{3/2}$ & 1$d_{5/2}$ & 0.245  & 0.210 \\ 
 $\pi$ & 2$s_{1/2}$ & 1$p_{3/2}$ & -0.136 & -0.131 & $\nu$ & 1$p_{1/2}$ & 7$d_{3/2}$ & 0.128  & 0.099  &       &            &            &        &       \\  
 $\pi$ & 1$p_{3/2}$ & 1$d_{5/2}$ & 0.337  & 0.322  & $\nu$ & 1$p_{1/2}$ & 8$d_{3/2}$ & 0.116  & 0.094  &       &            &            &        &       \\  
       &            &            &        &        & $\pi$ & 2$s_{1/2}$ & 1$p_{3/2}$ & -0.130 & -0.12  &       &            &            &        &       \\  
       &            &            &        &        & $\pi$ & 1$p_{3/2}$ & 1$d_{5/2}$ & 0.322  & 0.294  &       &            &            &        &       \\  
\hline
\end{tabular}
\caption{Main RPA components of the wavefunction of states associated with the soft dipole mode of $^{11}$Li, of excitation energy 
0.65, 1.21 and {2.00 MeV}  respectively. }
\end{table}
% n-inicial de cada l-j-tz
% 1s1/2 n: 1
% 1p3/2 n: 26
% 1p1/2 n: 50
% 1d5/2 n: 74
% 1d3/2 n: 98
% 1s1/2 p: 122
% 1p3/2 p: 147
% 1d5/2 p: 197
% e 1d3/2 p: 221

%GDR
\begin{table}[]
\begin{tabular}{lllll|lllll|lllll}
	% E = 24 MeV                                               % E = 25 MeV                                                % E = 26 MeV
      & i          & j          &$X_{ij}$&$Y_{ij}$ &       & i          & j          &$X_{ij}$&$Y_{ij}$&       & i           & j          &$X_{ij}$&$Y_{ij}$ \\
         \hline	 
$\nu$ & 1$s_{1/2}$ & 6$p_{3/2}$ & -0.138 & 0.004   & $\nu$ & 1$s_{1/2}$ & 8$p_{3/2}$ & -0.185 & 0.004  & $\nu$ & 14$s_{1/2}$ & 1$p_{3/2}$ & 0.175  & -0.003 \\  
$\nu$ & 1$s_{1/2}$ & 7$p_{3/2}$ & 0.176  & 0.004   & $\nu$ & 1$s_{1/2}$ & 9$p_{3/2}$ & 0.101  & 0.003  & $\nu$ & 15$s_{1/2}$ & 1$p_{1/2}$ & -0.687 & 0.002  \\ 
$\nu$ & 1$s_{1/2}$ & 6$p_{1/2}$ & 0.185  & -0.004  & $\nu$ & 1$s_{1/2}$ & 7$p_{1/2}$ & 0.248  & -0.004 & $\nu$ & 1$p_{3/2}$  & 8$d_{5/2}$ & -0.265 & 0.004  \\ 
$\nu$ & 1$s_{1/2}$ & 7$p_{1/2}$ & -0.155 & -0.004  & $\nu$ &12$s_{1/2}$ & 1$p_{3/2}$ & 0.118  & -0.007 & $\nu$ & 1$p_{3/2}$  &13$d_{3/2}$ & 0.406  & -0.002 \\ 
$\nu$ &11$s_{1/2}$ & 1$p_{3/2}$ & 0.134  & -0.007  & $\nu$ &13$s_{1/2}$ & 1$p_{3/2}$ & -0.386 & -0.006 & $\nu$ & 1$p_{1/2}$  &14$d_{3/2}$ & 0.326  & 0.003  \\ 
$\nu$ &12$s_{1/2}$ & 1$p_{3/2}$ & -0.216 & -0.006  & $\nu$ &14$s_{1/2}$ & 1$p_{1/2}$ & 0.119  & 0.003  & $\pi$ & 1$s_{1/2}$  & 3$p_{1/2}$ & -0.189 & 0.006  \\ 
$\nu$ & 1$p_{3/2}$ & 9$d_{5/2}$ & -0.119 & 0.015   & $\nu$ & 1$p_{3/2}$ &10$d_{5/2}$ & -0.104 & 0.013  & $\pi$ & 2$s_{1/2}$  & 1$p_{3/2}$ & -0.228 & 0.005  \\ 
$\nu$ & 1$p_{3/2}$ &10$d_{5/2}$ & -0.281 & 0.012   & $\nu$ & 1$p_{3/2}$ &11$d_{5/2}$ & -0.209 & 0.011  & $\pi$ & 1$p_{3/2}$  & 1$d_{5/2}$ & 0.109  & -0.010 \\
$\nu$ & 1$p_{3/2}$ &11$d_{5/2}$ & 0.303  & 0.010   & $\nu$ & 1$p_{3/2}$ &12$d_{5/2}$ & 0.464  & 0.009  &       &             &            &        &        \\  
$\nu$ & 1$p_{3/2}$ &10$d_{3/2}$ & 0.208  & -0.005  & $\nu$ & 1$p_{3/2}$ &11$d_{3/2}$ & 0.133  & -0.005 &       &             &            &        &        \\  
$\nu$ & 1$p_{1/2}$ &11$d_{3/2}$ & -0.384 & 0.010   & $\nu$ & 1$p_{3/2}$ &12$d_{3/2}$ & -0.133 & -0.004 &       &             &            &        &        \\  
$\nu$ & 1$p_{1/2}$ &12$d_{3/2}$ & 0.141  & 0.008   & $\nu$ & 1$p_{1/2}$ &13$d_{3/2}$ & -0.280 & 0.009  &       &             &            &        &        \\  
$\pi$ & 1$s_{1/2}$ & 1$p_{3/2}$ & 0.570  & -0.004  & $\nu$ & 1$p_{1/2}$ &14$d_{3/2}$ & 0.158  & 0.007  &       &             &            &        &        \\  
      &            &            &        &         & $\pi$ & 1$p_{3/2}$ & 1$d_{5/2}$ & 0.459  & 0.004  &       &             &            &        &        \\  
\hline                                                                                                   
\end{tabular}                                                                                            
\caption{Main RPA components of the wavefunctions of  states associated with the GDR of $^{11}$Li, of excitation energy 18.61, 22.13 and 27.26 MeV respectively.}                                                   
\end{table}                                                                                              
            
\clearpage
\bibliographystyle{ieeetr}                                                                               
\bibliography{./pygmy_enr_a}{}                                                                               
 
 \end{document}